\documentclass[useAMS,usenatbib,usegraphicx]{mn2e}
 
\usepackage{natbib}
\usepackage{epsfig}
\usepackage{amssymb}
\usepackage{amsfonts}
\usepackage{amsmath}
\usepackage{color}
 
\input epsf

\begin{document}

\title[Photometric redshifts for 1.5M LRGs]{A Comparison of Six Photometric Redshift Methods Applied to 1.5 Million Luminous Red Galaxies}

\author[F. B. Abdalla et al.]{F. B. Abdalla$^1$ \thanks{E-mail: fba@star.ucl.ac.uk},
  M. Banerji$^1$ \thanks{E-mail: mbanerji@star.ucl.ac.uk}, O.Lahav$^1$ \thanks{E-mail: lahav@star.ucl.ac.uk}, V. Rashkov$^{2}$\\ \\ 
  $^1$ Department of Physics \& Astronomy, University College London, Gower
  Street, London, WC1E 6BT, UK \\ 
  $^2$ Department of Astrophysical Sciences, Princeton University, Princeton, New Jersey 08544, USA\\
}
\maketitle

\begin{abstract}

We present an updated version of 
MegaZ-LRG \citep{2007MNRAS.375...68C} with photometric redshifts derived with
the neural network method, ANNz as well as five other publicly available photo-z codes (HyperZ,
SDSS, Le PHARE, BPZ and ZEBRA) for $\sim$1.5 million Luminous Red Galaxies (LRGs) in SDSS DR6.  
This allows us to identify 
how reliable codes are relative to each other if used as described 
in their public release. We compare and contrast the
relative merits of each code using 
$\sim$13000 spectroscopic redshifts from the 2SLAQ sample.
We find that the performance of each code depends on the figure of merit used to assess it. 
As expected, the availability of a 
complete training set means that the training method performs best 
in the intermediate redshift bins where there are plenty of training 
objects. Codes
such as Le PHARE, which use new observed templates perform best in the lower 
redshift bins. 
All codes produce 
reasonable photometric redshifts, the 1-$\sigma$ 
scatters ranging from 0.057 to 
0.097 if averaged over the entire redshift range.
We also perform tests to check whether a training set from a small region of 
the sky such as 2SLAQ produces biases if used to train over a larger area 
of the sky. We
conclude that this is not likely to be a problem for future wide-field surveys. The complete photometric
redshift catalogue including redshift estimates and errors on these from all six methods can be found at www.star.ucl.ac.uk/$\sim$mbanerji/MegaZLRGDR6/megaz.html

\end{abstract}


\section{Introduction}

Photometric redshifts will be one of the key ingredients for us to improve 
our understanding of the Universe in the decade to come. Up to date, 
galaxy large scale structure surveys relied mainly on spectroscopic redshifts 
to produce high precision power spectrum measurements of the galaxy 
distribution \citep[e.g.][]{2005MNRAS.362..505C,2007ApJ...657..645P}.
Combined with CMB experiments these surveys have provided evidence 
that the Universe is flat and is likely to be dominated by a dark energy 
component \citep{2008arXiv0803.0547K}.

\begin{figure*}
\begin{center}
\includegraphics[width=15.5cm,angle=0]{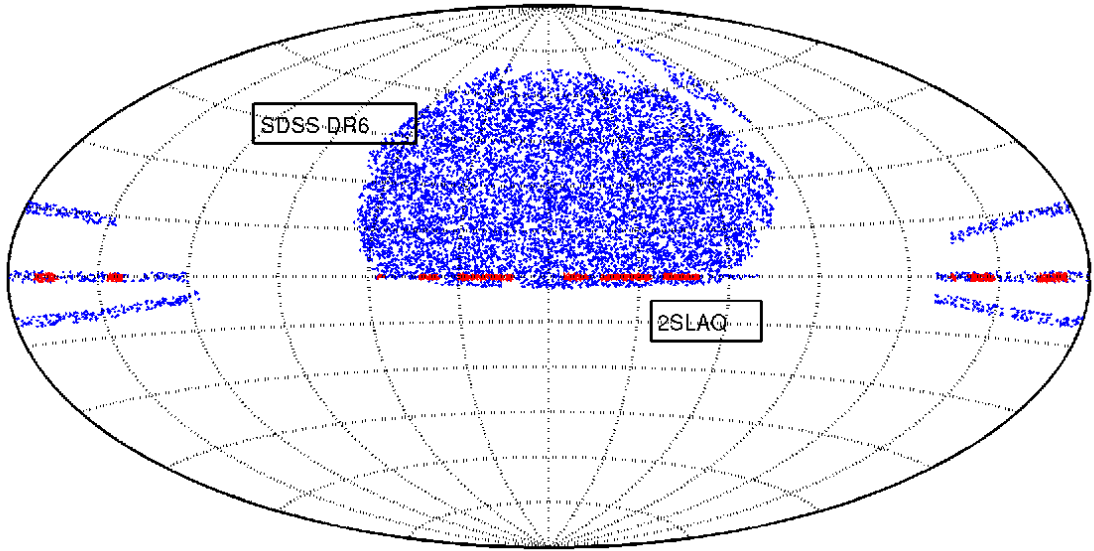}
\caption{Map of the MegaZ-LRG sample (blue) covering the SDSS DR6 area
as well as the 2SLAQ sample (red). For clarity 
only a random subsample of galaxies have been plotted.}
\end{center}
\end{figure*}

However having a considerable step up in the size of the spectroscopic 
surveys will be a hard task to achieve for technical reasons. Several 
Multi-fibre optical spectrographs are currently being built (FMOS) 
\citep{2006SPIE.6269E.136D} or being designed (WFMOS), but it is unlikely 
that they will be able 
to survey a considerable part of the sky. 
On the other hand radio interferometers 
may be able to perform spectroscopic surveys of 
the sky reasonably quickly
\citep{2004NewAR..48.1063B,2005MNRAS.360...27A}
but the timescale for the technical advances to allow for this 
will be relatively long.

The alternative to a full spectroscopic survey 
is to obtain multi-colour images of the 
sky and perform photometric redshift estimates for the galaxies we have 
available \citep[e.g.][]{2003AJ....125..580C}. In a pilot 
study with high redshift Luminous Red Galaxies (LRGs) it has been shown 
\citep{2006astro.ph..5302P,2007MNRAS.374.1527B}
that producing large scale measurements with 
photometric redshifts is possible and competitive with a smaller 
spectroscopic redshift survey. Using the same dataset
\citet{2008MNRAS.385.1257B} have shown that photometric redshifts can
also be used to study small scale halo model signatures.

On the other hand there are many caveats of photometric redshifts that 
have to be assessed in order for us to be completely confident that these 
measurements are reliable to the level of systematics that we expect in 
future surveys. For instance \citet{2007MNRAS.374.1527B} have performed a 
detailed study of 
whether star-galaxy separation influences the cosmological measurements 
given that the LRGs that have been selected are contaminated at the per cent
level by M-type stars which have similar colours. They have 
also assessed whether there is a significant contamination from dust 
corrections in the galaxy, by obtaining estimates of the power spectrum 
in different regions of obscuration in the sky.

We extend this analysis concentrating on the level of 
systematic effects that is introduced by the use of different photometric 
redshift techniques. We have selected the same sample as was selected 
in the MegaZ-LRG catalogue \citep{2007MNRAS.375...68C} 
and used several different photometric redshift 
techniques on the same galaxies available from the literature, 
including artificial neural networks, 
template fitting techniques and Bayesian techniques. We note here that 
LRGs have well defined 4000 $\AA$ break, hence this strong feature
makes photometric redshift 
estimation an easier task. Here all codes compared produce 
reasonable photometric redshifts and we are comparing more subtle 
differences between codes.

In Sec.\ref{sec:data} we describe the MegaZ-LRG data used. 
In Sec.\ref{sec:methods} we describe all the methods we have used to 
estimate the photometric redshifts for the LRG sample. 
In Sec.\ref{sec:results} we compare different statistics for the 
different photo-z results.
We perform an analysis to check for gradients across the sky which could 
arise from training sets if they only belong to a small area of the sky 
in Sec.\ref{sec:harmonics} and we present the catalogue in 
Sec.\ref{sec:catalogue}. Our conclusions are drawn in Sec.\ref{sec:concl}.

\section{Data}
\label{sec:data}

We use galaxy photometry in a DR6 equivalent to the MegaZ-LRG catalogue, a
photometric-redshift catalogue of Luminous Red Galaxies based on the
imaging component of the SDSS 4th Data Release.  The construction of
this catalogue follows the same prescription as in \citet{2007MNRAS.375...68C}.
Here we only outline briefly the description of the catalogue. 
For details on the construction of this catalogue see 
\citet{2007MNRAS.375...68C}. 

\begin{table*}
\begin{center}
\begin{tabular}{|l|l|l|l|}
\hline
\textbf{Code}
&\textbf{Authors}
&\textbf{Method}
&\textbf{Web link}
\\
\hline
HyperZ &  Bozonella et al.  &  Template & http://webast.ast.obs-mip.fr/hyperz/   \\
SDSS template & SDSS pipeline & LRG Template & N/A code obtained from N. Padmanabhan\\
BPZ  & Benitez &  Template + Bayesian priors & http://acs.pha.jhu.edu/$\sim$txitxo/bpzdoc.html  \\
ANNz & Collister \& Lahav  & Neural Networks &http://zuserver2.star.ucl.ac.uk/$\sim$lahav/annz.html \\

ZEBRA & Feldmann et al. & Template, Bayesian, Hybrid  & www.exp-astro.phys.ethz.ch/ZEBRA \\
Kcorrect & Blanton & Model templates &  http://cosmo.nyu.edu/blanton/kcorrect/ \\
Le PHARE & Arnouts \& Ilbert & Template & www.oamp.fr/people/arnouts/LE\_PHARE.html\\
EAZY & Brammer at al. & Template & www.astro.yale.edu/eazy/ \\
LRT Libraries & Assef et al. & Template & http://www.astronomy.ohio-state.edu/$\sim$rjassef/lrt/\\ 
\hline
\end{tabular}
\end{center}
\caption{\label{tab:codes} Publicly available software packages for 
photo-z estimation, to date and to our knowledge. 
In this work we have used six representative codes from this table, namely HyperZ, SDSS, BPZ, ANNz, ZEBRA and Le PHARE.}
\end{table*}

\subsection{Selection criteria}

The MegaZ-LRG catalogue is selected from the SDSS imaging database 
using a series of colour and magnitude cuts \citep{2007MNRAS.375...68C}
which were designed to match the selection criteria of the 2dF-SDSS LRG 
and Quasar (2SLAQ) survey \citep{2006MNRAS.372..425C}. 
2SLAQ is a spectroscopic follow-up combining the SDSS photometric 
survey and the spectroscopy from the Two-degree Field (2dF) instrument 
of the Anglo-Australian Telescope (AAT).

The spectroscopic redshifts available from 2SLAQ were
used to train and test the photometric redshift code, which we then
applied to the entire set of LRGs selected from the SDSS imaging
database. Around $13{,}000$
objects in selected fields of the SDSS equatorial stripe (at
declination $\delta \approx 0^\circ$) were available.
The 2SLAQ survey demonstrated that these selection
criteria are $\approx 95\%$ efficient in the identification of
intermediate-redshift LRGs.  The most significant contaminant,
accounting for virtually all of the remaining $\approx 5\%$ of
objects, is M-type stars.

The 2SLAQ selection criteria fluctuated a little at the beginning of
the survey.  Specifically, the faint limit of the $i$-band magnitude
$i_{\rm deV}$, and the minimum value of $d_{\perp}$ (a colour
variable used to select LRGs), were varied slightly.  For the majority
of the 2SLAQ survey, the criteria $i_{\rm deV} \le 19.8$ and $d_{\perp} 
\ge 0.55$ were used. For further details on this see 
\citep{2006MNRAS.372..425C}.

We note that our training sub-sample is extrapolated in sky position.
The 2SLAQ targets lie exclusively in the equatorial stripe at
declination $\delta \approx 0^\circ$, so may not fully trace effects
such as dust extinction which depend on sky co-ordinate. One of the important 
aims of this study is to assess how much this sky extrapolation biases the
final photo-z measurements with a training set method.

\section{Photometric redshifts estimators}
\label{sec:methods}

This section describes how we obtained the photometric redshifts for 5482 
galaxies in the 2SLAQ sample. We have subdivided the galaxies from 2SLAQ into
a training sample and a testing sample. The training sample was used to train 
the training part of the codes presented here. 
The rest of the sample was chosen to be 5482
so that enough galaxies were left in the training sample. These galaxies were 
randomly chosen. Only these galaxies
were used to test the codes final accuracy, hence there being less galaxies 
than the 13000 galaxies available to test the codes.

We have used several different codes
in the literature to provide photometric redshift estimates as well as
redshift errors for a subsection of the 2SLAQ sample. The rest of the sample 
was used by some of the methods as a training set and therefore we do 
not use those galaxies in the comparison as this might introduce biases in our 
study.

We emphasise here that the comparison we are undertaking 
is a high level comparison; i.e. we are comparing end products without 
decomposing the problem into smaller parts in order to potentially 
assess where discrepancies are arising, in other words 
comparing codes as black boxes. 
Therefore the comparison is 
a comparison of the ensemble of codes plus galaxy libraries used with 
each code. We argue that this is a valid comparison as this is what a 
naive user of these publicly available codes would get should they choose 
one code rather than another. We also argue that a full analysis
is needed to have the highest level of confidence in photometric redshifts
and believe that this is being done by the Photometric accuracy testing 
program (PHAT) 
collaboration\footnote{http://www.astro.caltech.edu/twiki\_phat/bin/view/}.

\subsection{Methodology: codes considered in this work}
\label{sec:codes}

We give here a brief description of the codes we have used 
in this work. For a more general description of photo-z methods
we refer the reader to \citet{PhotozBook} and \citet{2008arXiv0811.2600B}.

\subsubsection{SDSS template fitting code.}
\label{sed::sdss_code}

The template-fitting technique in photometric redshift estimation 
is a $\chi^{2}$ fit between the data and a given set of templates
for those galaxies. For the purpose of redshift estimation, 
the galaxy templates usually come from stellar population synthesis
models \citep[e.g.][]{1997A&A...326..950F,2003MNRAS.344.1000B}. 
A linear combination of templates is used. 
The coefficients $c_i$ of the templates are the free parameters for the 
minimisation. 
We note $\Psi^{i}(z)$ the set of templates observed at redshift z and 
$f_{\alpha}$ the observed flux in filter $\alpha$ with error of 
$\sigma_{\alpha}$. The photometric redshift is found via
$\chi^{2}$ 

\begin{equation}
\chi^{2}(c_i,z) = \sum_{\alpha}\left({f_{\alpha} - R_{\alpha}(\sum_{i}c_{i}
\Psi^{i}(z)) \over \sigma_{\alpha}}\right)^{2}
\end{equation}

\noindent where $R_{\alpha}(\Psi)$ is the flux of spectrum 
$\Psi$ through filter $\alpha$. 

The greatest disadvantage of this method (which also applies to all the other
template fitting techniques presented here) is the potential mismatch 
between the templates used for the fitting and the properties of the 
sample of galaxies for which one wants to estimate the redshifts. 
A hybrid method can be used, in which in order to calibrate the templates 
to a better representation of the studied galaxy sample one would 
use a training set with known spectroscopic redshifts and
similar properties to the galaxies whose redshifts need to be estimated. 
The SDSS code used here applies a hybrid method to the LRG sample 
using a modified elliptical galaxy template, 
adjusted to represent an LRG spectrum after three iterations of correction.
Given that early type galaxies evolve passively, only one template is used in
the code.

\subsubsection{HyperZ}

HyperZ \citep{2000A&A...363..476B} was the first publicly 
available photo-z code and has consequently 
been widely used in the literature for photometric redshift estimation. 
It is a simple template fitting code that can be used in conjunction with 
two sets of basis SEDs, namely the observed \citet{1980ApJS...43..393C} 
templates (CWW hereafter) or the synthetically generated 
\citet{2003MNRAS.344.1000B} templates (BC hereafter). HyperZ takes as its 
inputs the photometric catalogue of galaxies with magnitudes and 
errors on magnitudes through the different filters specified in the 
filter set, as well as a list of spectral templates to be used in the 
$\chi^2$ minimisation.  Various different reddening laws can also be 
implemented in order to account for the effect of interstellar dust on 
the shape of the SED. The damping of the Ly$\alpha$ forest increasing with 
redshift is modelled according to \citet{1995ApJ...441...18M}. We have 
experimented with a variety of different basis template sets including 
the four CWW templates and interpolations between them as well as the 8 
synthetic BC templates. We find the BC templates to produce considerably 
better photo-z's than the CWW and interpolated CWW template sets. 
In order to demonstrate the effects of using two different template sets 
to calculate photometric redshifts with the same code, we present results 
obtained using both the four CWW templates roughly corresponding to types 
E,Sbc,Scd and Im and the eight BC templates roughly corresponding to types 
Single Burst, E, S0, Sa, Sb, Sc, Sd and Im. 

We have also considered the photo-z outputs with no correction for 
galactic reddening and with a Calzetti reddening law 
\citep{1994ApJ...429..582C} applied to the 
templates for all the template sets considered. In all cases we find that 
including an extinction correction slightly worsens the photometric 
redshift estimate. Our final HyperZ outputs therefore make no correction 
for the galactic extinction.  

We used magnitudes in all five SDSS optical bands even though the 
photometric uncertainties in the $u$-band are large and therefore would 
contribute to a larger scatter in the photo-z estimate. We have checked 
that removing the $u$-band data does in fact worsen the photo-z estimate.

We note here that other template-based methods are also available for photo-z estimation
such as ImpZ (private communication: M. Rowan-Robinson), k-correct \citep{2007AJ....133..734B}, EAZY \citep{2008ApJ...686.1503B}  and the LRT Libraries \citep{2008ApJ...676..286A}. 
These have not been presented in this comparison.

\subsubsection{ANNz}
\label{sec::annz_code}

When a representative training set is available training methods become 
a viable option to use instead of template-fitting methods. 
The basic principle of training methods is the derivation of a 
parameterisation of redshift through the magnitudes of the galaxies in 
a training set. This parameterisation is then applied to galaxies for 
which no spectroscopy is available, yielding an estimate of the 
photometric redshift. One of the training methods used here is 
Artificial Neural Networks \citep{2004PASP..116..345C}. 
Neural networks have been used for 
estimation of photo-z in data \citep{2007MNRAS.375...68C} 
as well as forecasts 
of photometric redshifts for future data 
\citep{2008MNRAS.386.1219B,2008MNRAS.387..969A}.
An artificial neural network is made up of several layers, each consisting 
of a number of nodes. The first layer receives the galaxy magnitudes as
inputs, while the last layer outputs the estimated photometric redshift. 
The layers in between could consist of any number of nodes each. 
The nodes are interconnected so that a node in a given layer is connected to 
all nodes in the adjacent layers, every connection carrying a weight 
$w_{ij}$, where {\it i} and {\it j} describe the two nodes. 
Each node {\it i} is assigned a value $u_i$ and an activation function 
$g_i(u_i)$

\begin{equation}
g_i(u_i) = {1 \over 1 + exp(- u_i)}
\end{equation}

\noindent  is evaluated.

The value of a subsequent node {\it j} is then calculated as the 
summation of the weighted values of the activation functions of all 
nodes {\it i} pointing to it:

\begin{equation}
u_j = \sum_i w_{ij}g_i(u_i).
\end{equation}

When a network is trained the weights of all node connections
are determined by minimising a cost function {\it E} evaluated on the 
training set of galaxies where

\begin{equation}
E = \sum_k (z_{phot}(w, m_k) - z_{spec,k})^2
\end{equation}

\noindent and photometric input of $m_k$ for galaxy {\it k} from the training 
set is $z_{phot}(w, m_k)$, and the spectroscopic redshift of the galaxy 
is $z_{spec,k}$.

To avoid an over-fitting, every network is tested on a validation set of 
galaxies, whose spectroscopic redshifts are also known. 
The network with lowest value of {\it E} as calculated on the 
validation set is selected and the photometric sample is run through 
it for redshift estimation \citep{2004PASP..116..345C}. 

The artificial neural networks used in {\it ANNz} can be described as follows: 
$N_{in}:N_1:N_2:...:N_{out}$, where $N_{in}$ and $N_{out}$ are respectively 
the number of input and output parameters, while $N_i$ is the number of 
nodes in the $i^{th}$ intermediate layer. In the case of photometric 
redshift estimation using MegaZ-LRG, a network of the form 5:10:10:1 was used,
this was found empirically to be optimal \citep{2003MNRAS.339.1195F,2007MNRAS.375...68C}.

\begin{figure*}
\begin{center}
\begin{minipage}[c]{1.00\textwidth} 
\centering 
\begin{tabular}{cc}
\includegraphics[width=8.5cm,height=5.5cm,angle=0]{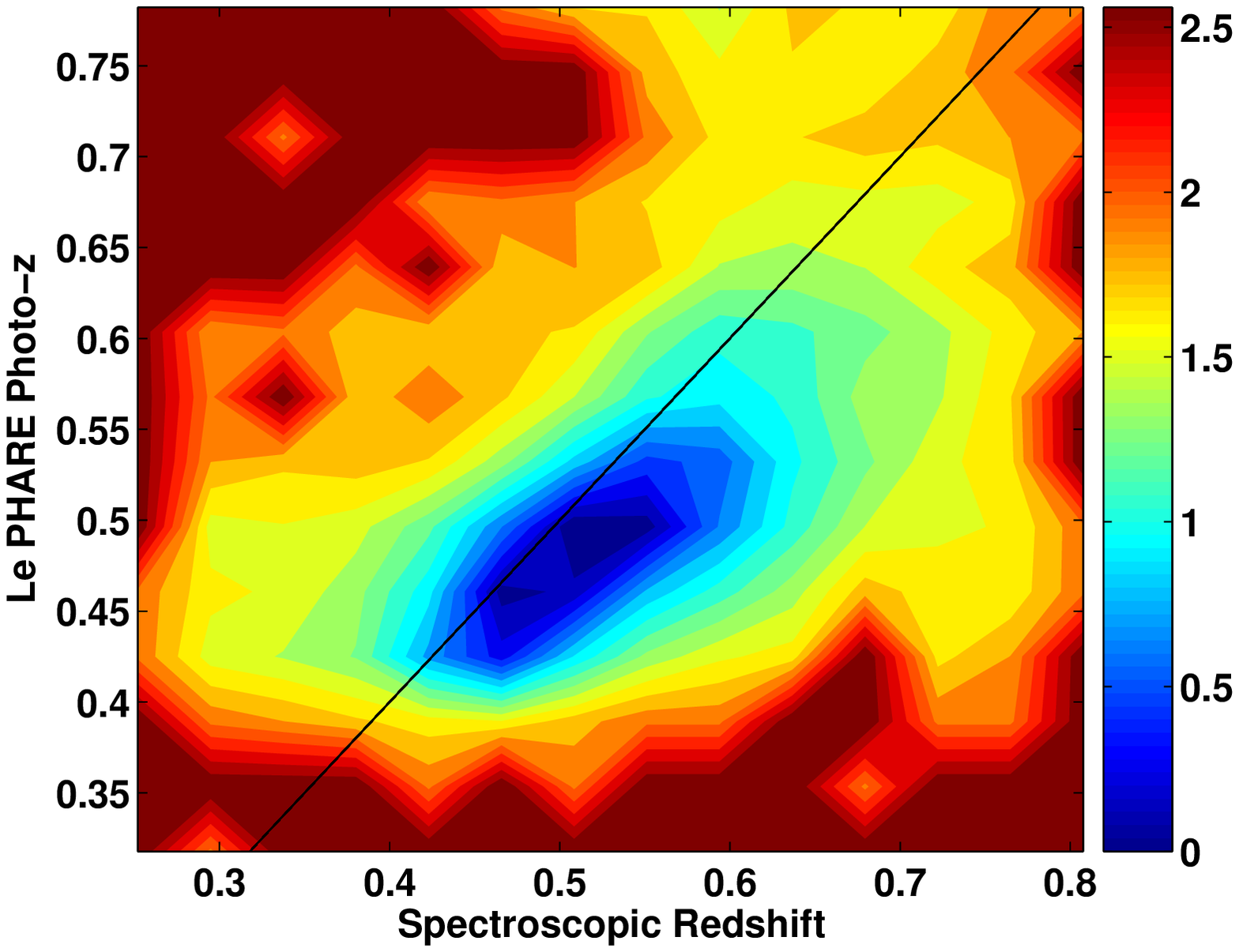} 
\includegraphics[width=8.5cm,height=5.5cm,angle=0]{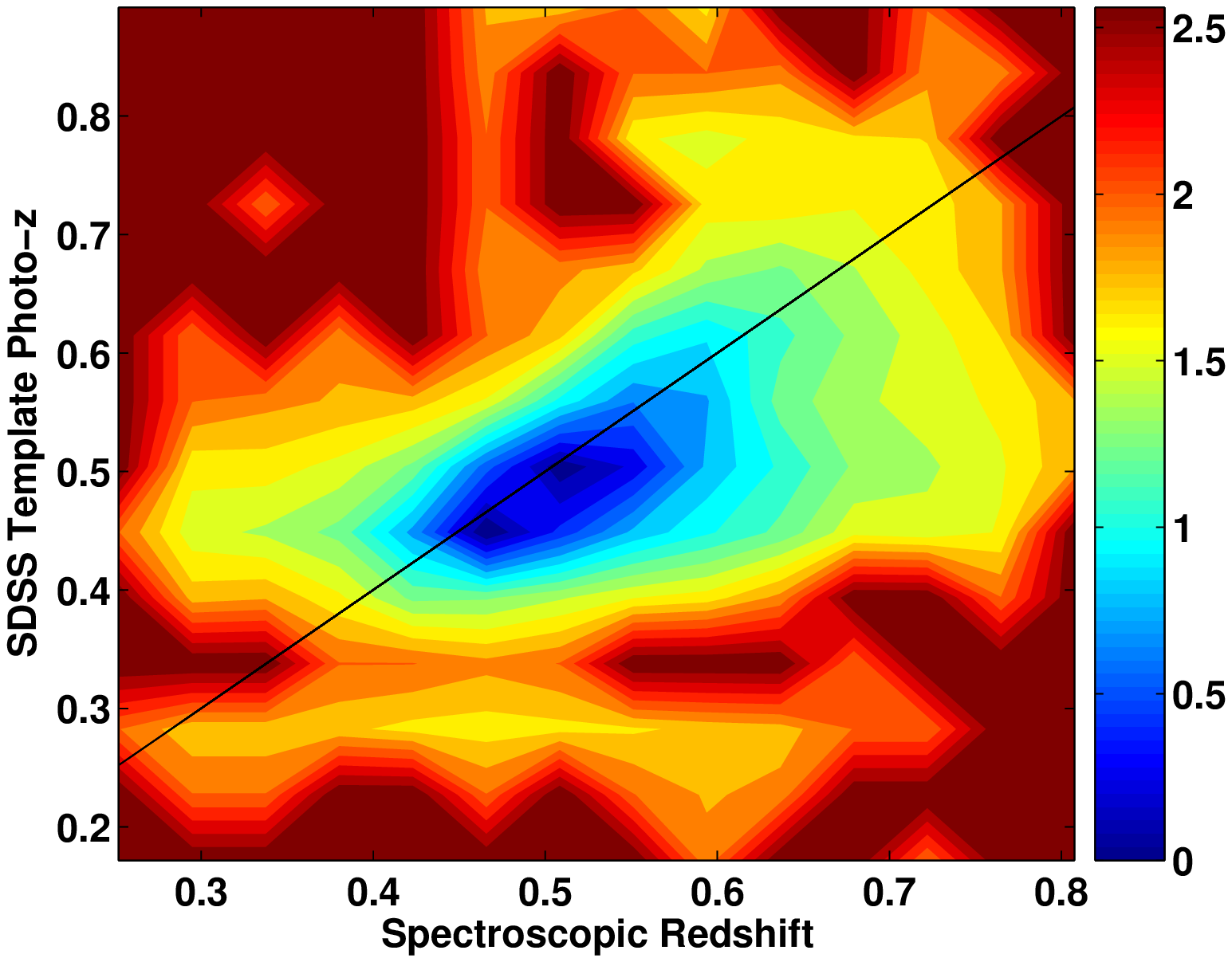} \\
\end{tabular}
\begin{tabular}{cc}
\includegraphics[width=8.5cm,height=5.5cm,angle=0]{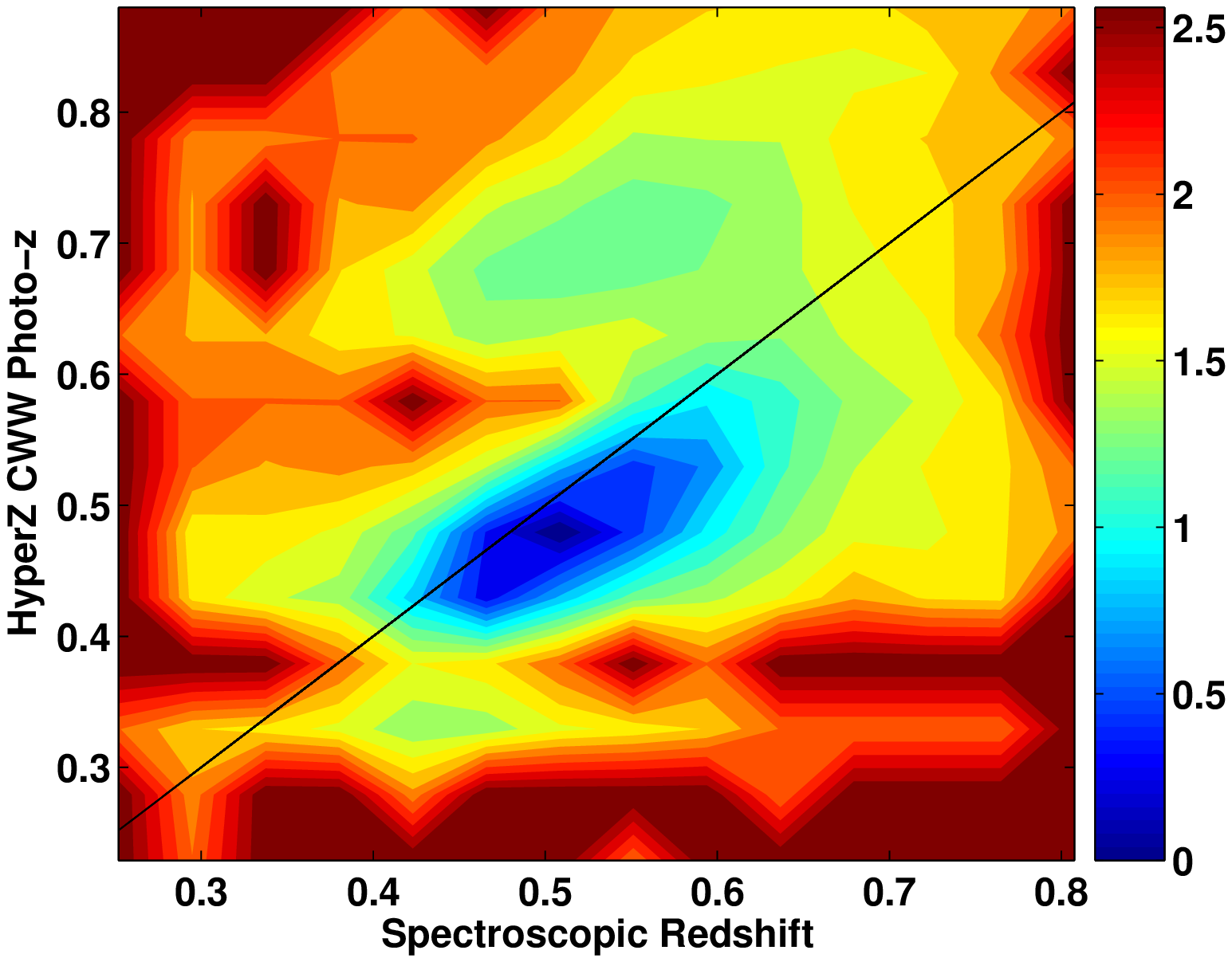}  
\includegraphics[width=8.5cm,height=5.5cm,angle=0]{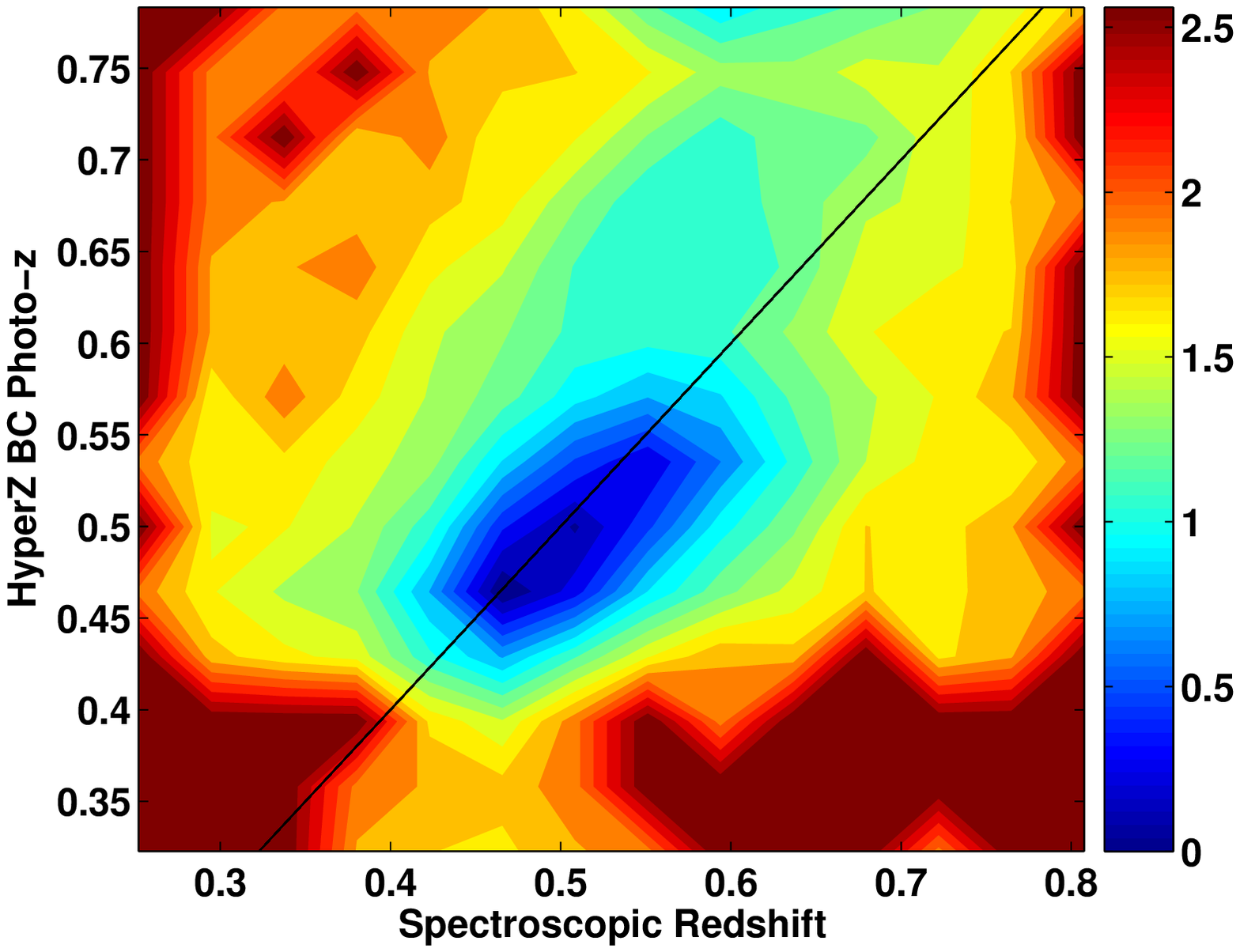} \\
\end{tabular}
\end{minipage}
\begin{minipage}[c]{1.00\textwidth} 
\centering 
\begin{tabular}{cc}
\includegraphics[width=8.5cm,height=5.5cm,angle=0]{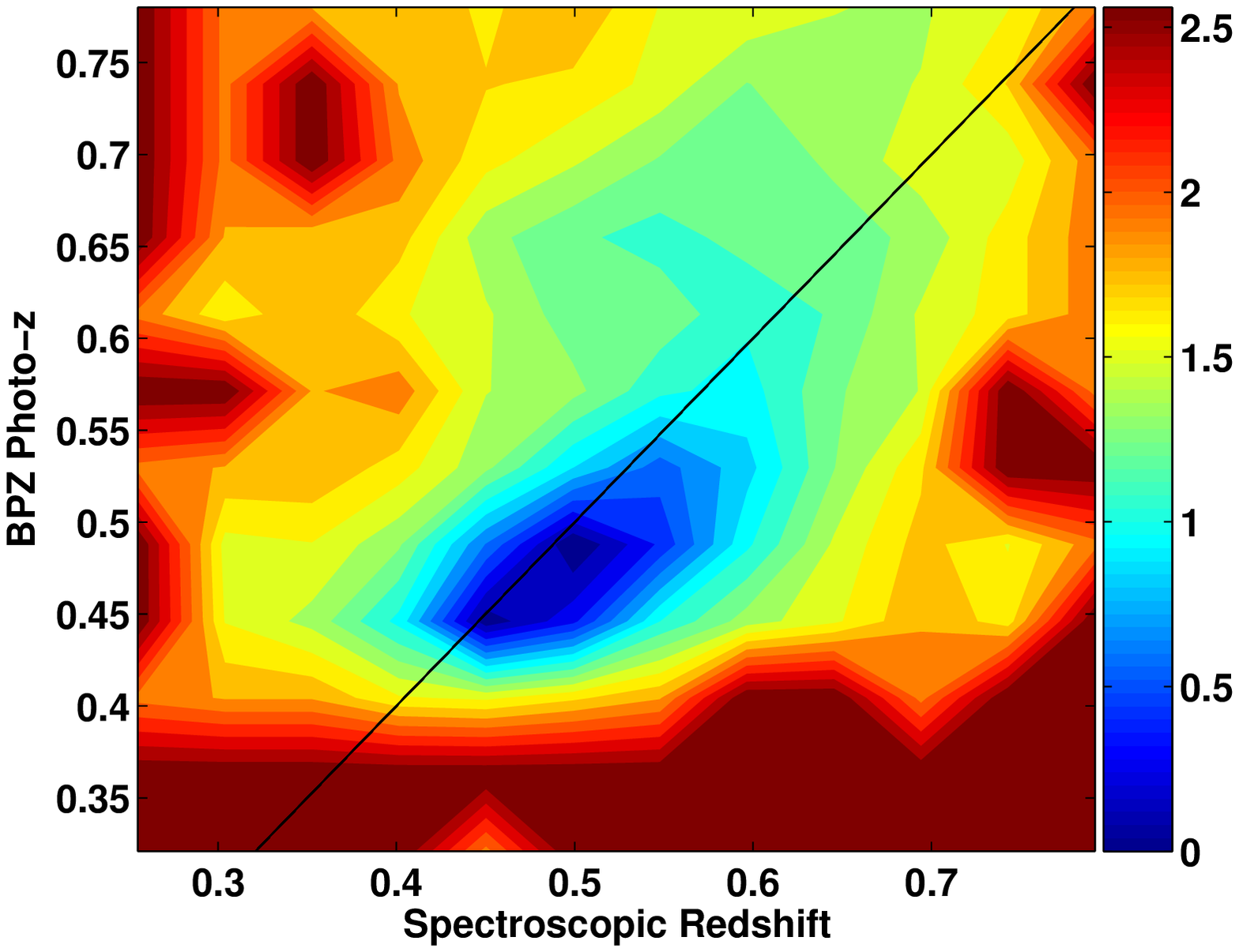}
\includegraphics[width=8.5cm,height=5.5cm,angle=0]{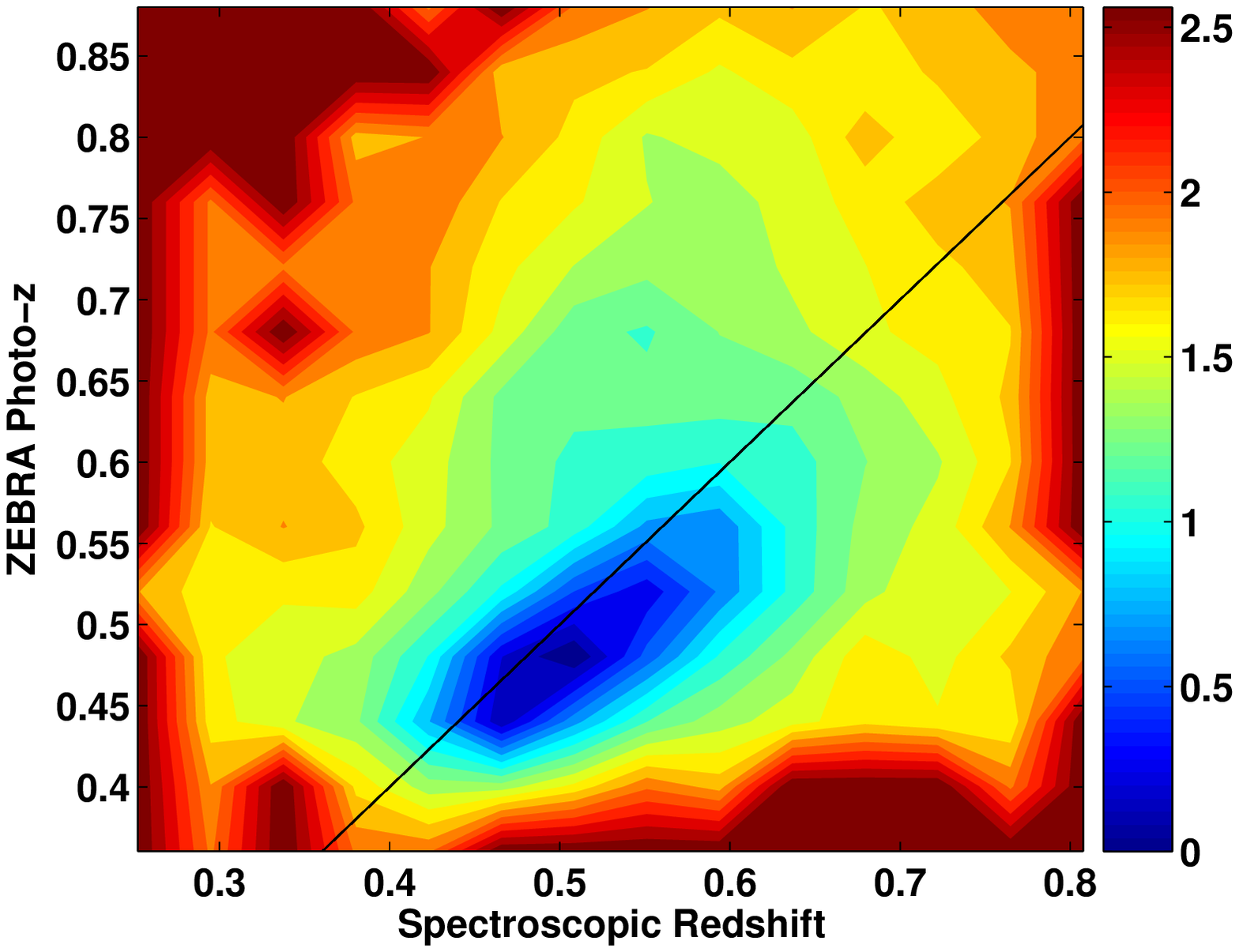} \\
\end{tabular}
\centering 
\begin{tabular}{cc}
\includegraphics[width=8.5cm,height=5.5cm,angle=0]{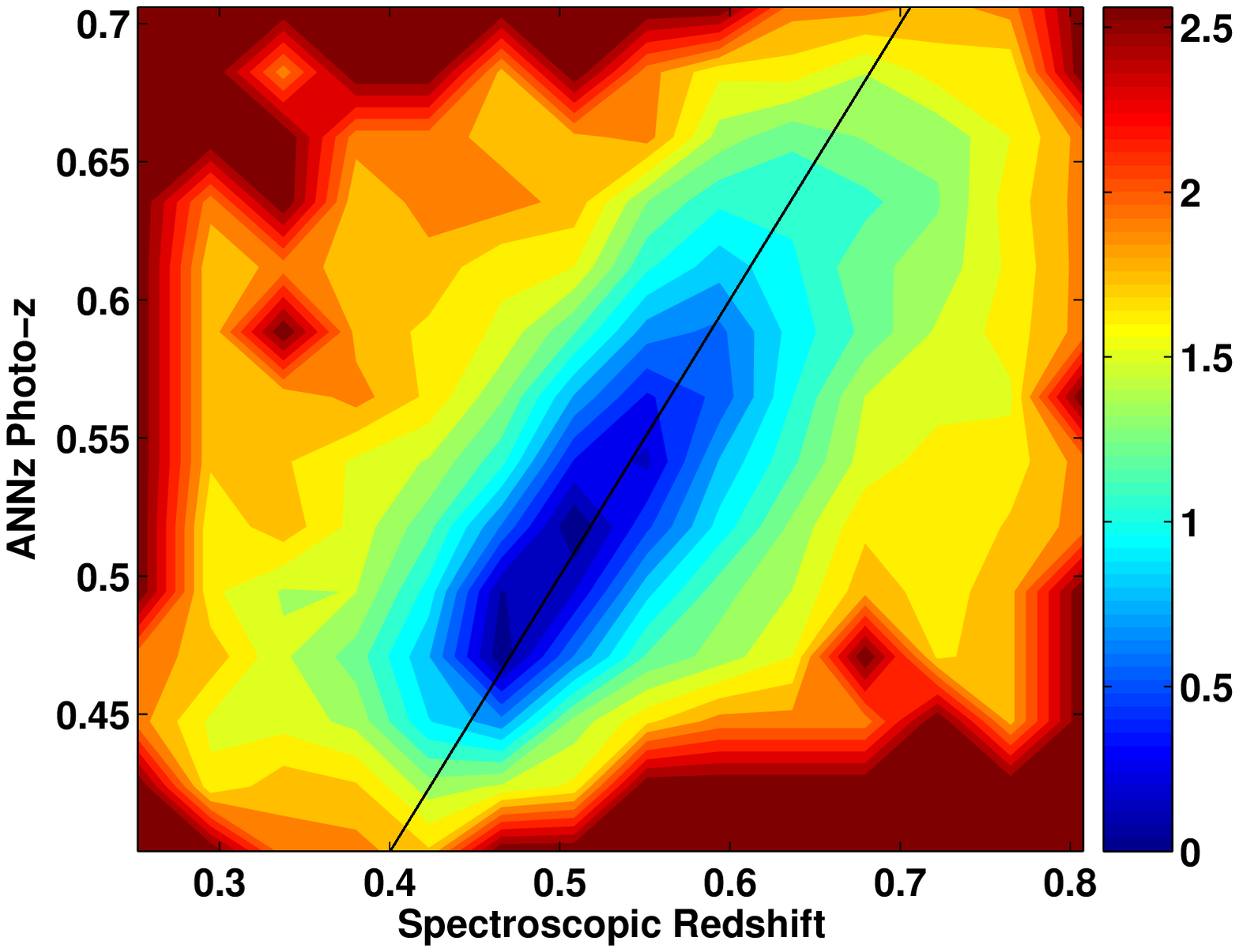} 
\end{tabular}
\end{minipage}
\caption{Density plots of spectroscopic versus photometric redshift for each of the public photo-z codes described in $\S$ \ref{sec:codes}.  The plots are colour-coded and the scale is exponential. A colour difference of one is equivalent to the density being decreased by a factor of $e$. The solid black lines show where the spectroscopic redshift is equal to the photometric redshift.
\label{fig:density}}
\end{center}
\end{figure*}

\begin{figure*}
\begin{center}
\begin{minipage}[c]{1.00\textwidth}
\centering 
\includegraphics[width=7.5cm,angle=0]{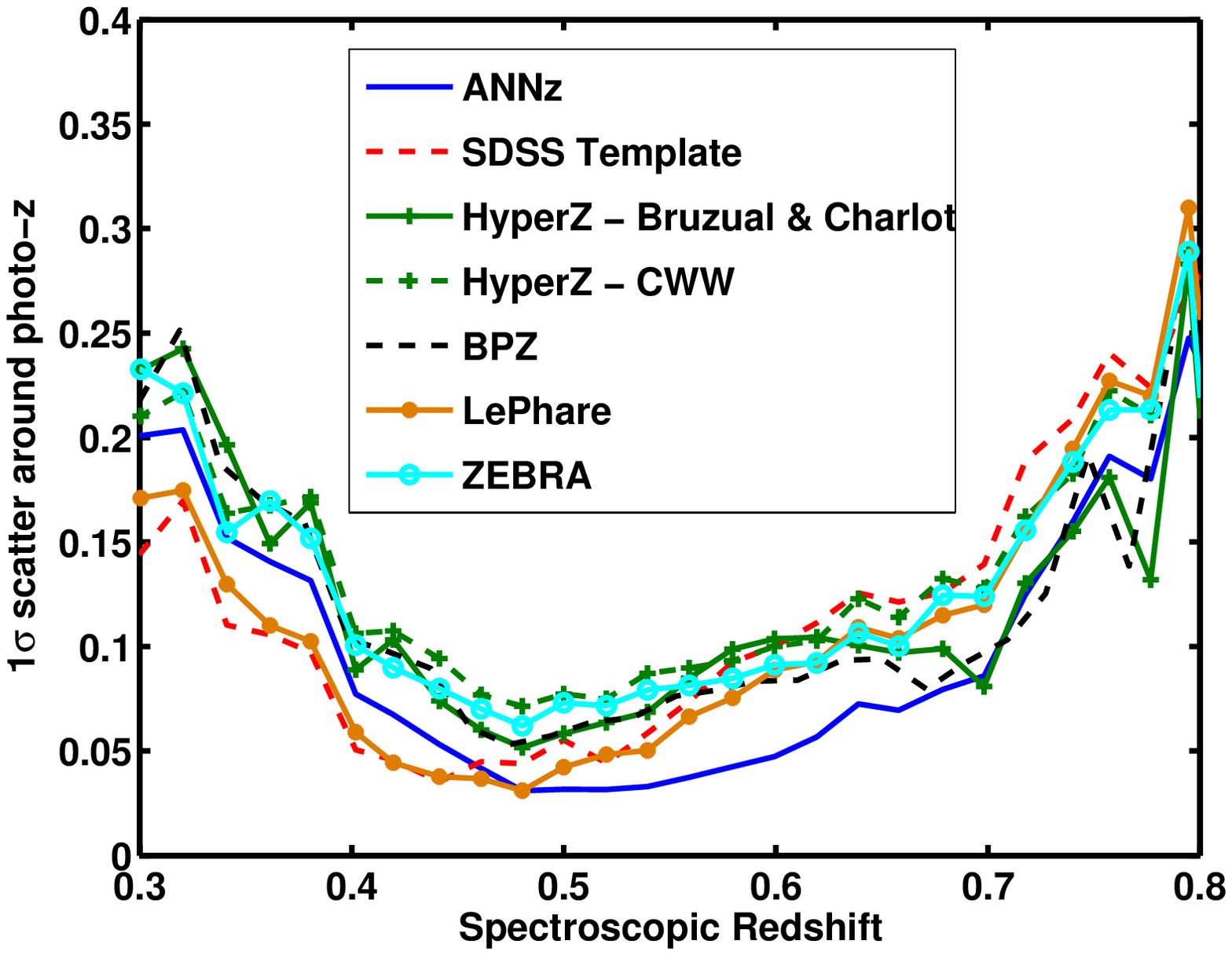}
\includegraphics[width=7.5cm,angle=0]{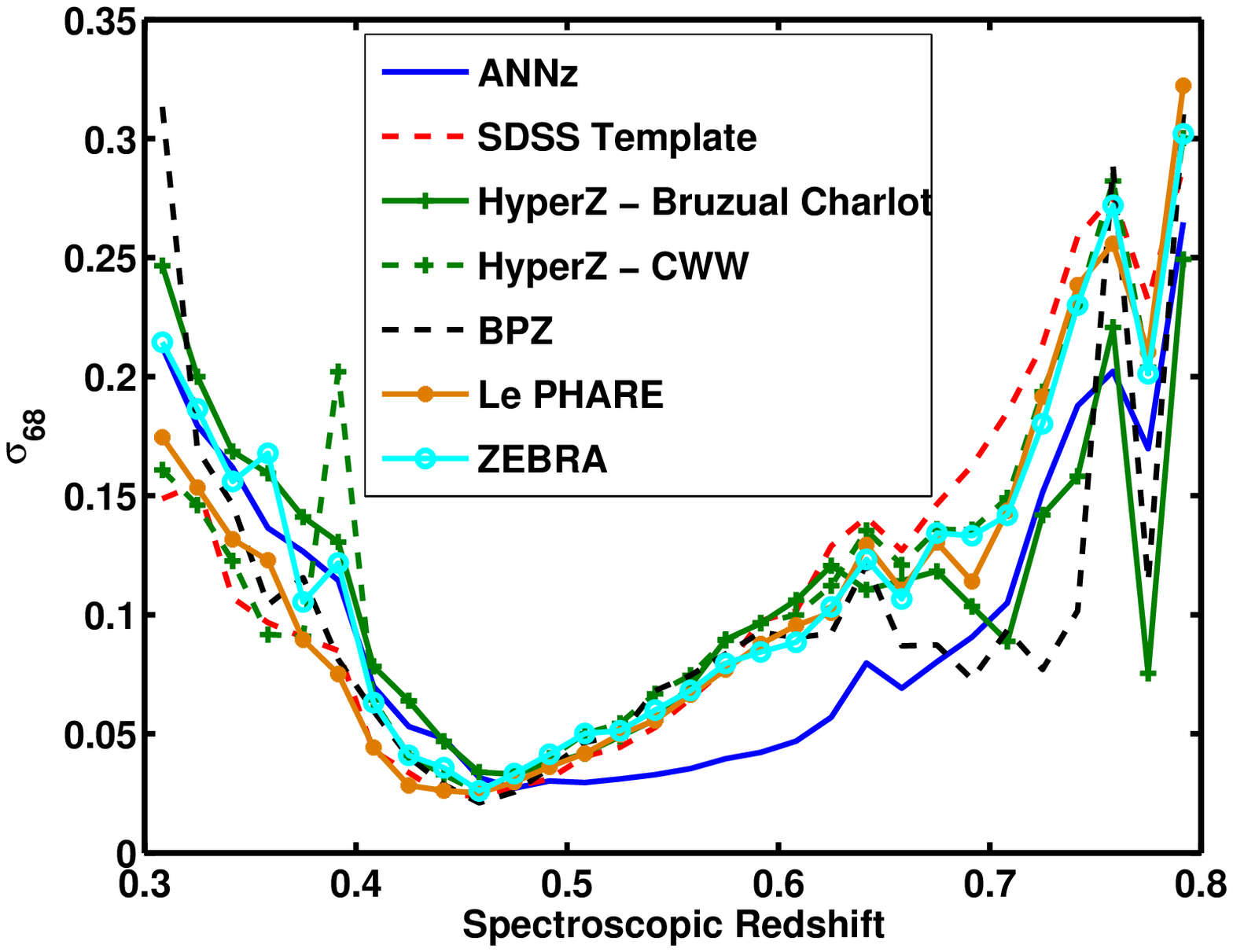}
\end{minipage}
\caption{$1\sigma$ scatter on the photometric redshift around the true spectroscopic redshift defined as per Eq.\ref{eq:scatter1} for each of the public photo-z codes described in $\S$ \ref{sec:codes} in the left-hand panel and $\sigma_{68}$ as a function of the spectroscopic redshift for each of the public photo-z codes described in $\S$ \ref{sec:codes} in the right-hand panel.}
\label{fig:scatter1}
\end{center}
\end{figure*}

\begin{figure*}
\begin{center}
\begin{minipage}[c]{1.00\textwidth}
\centering 
\includegraphics[width=7.5cm,angle=0]{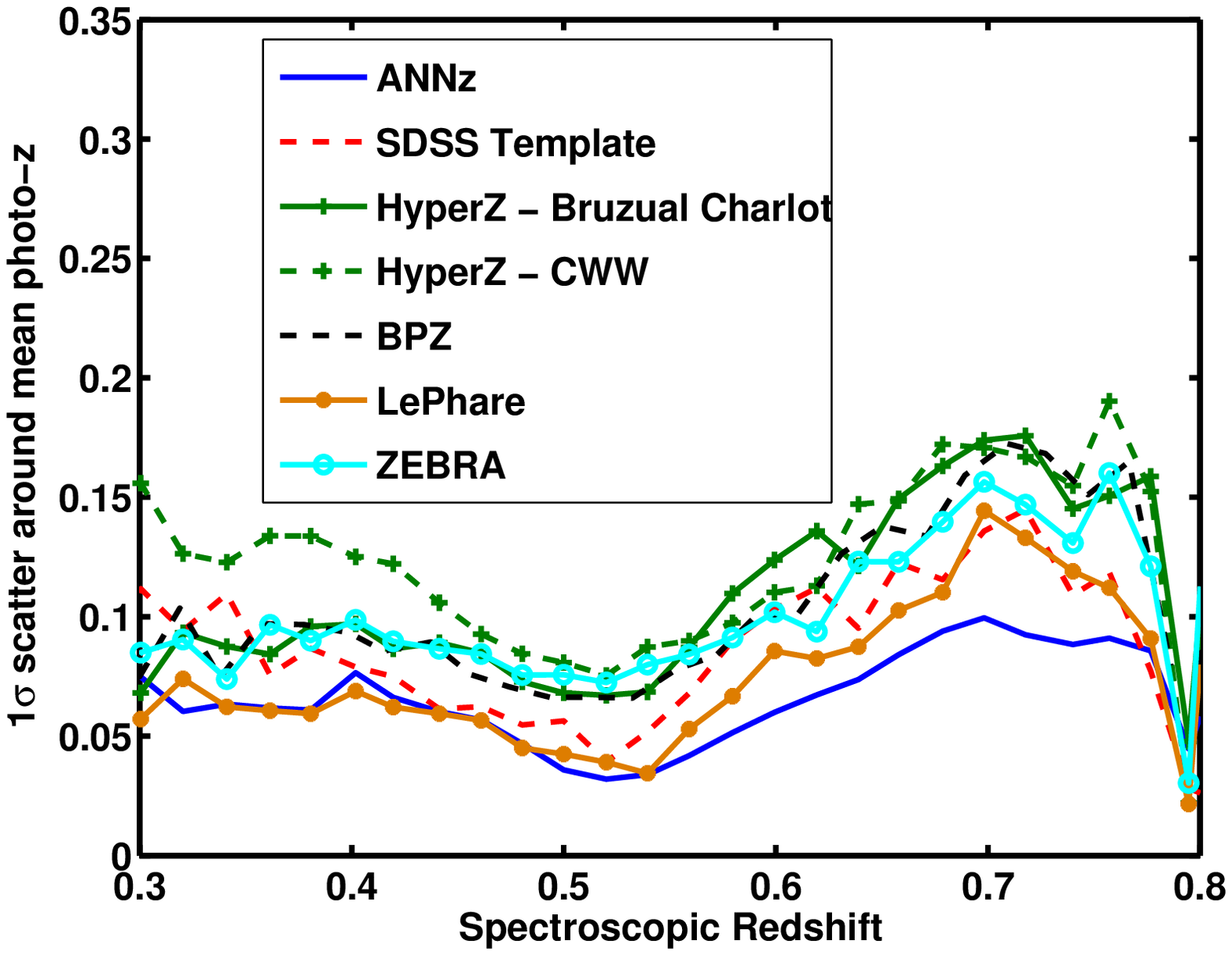}
\includegraphics[width=7.5cm,angle=0]{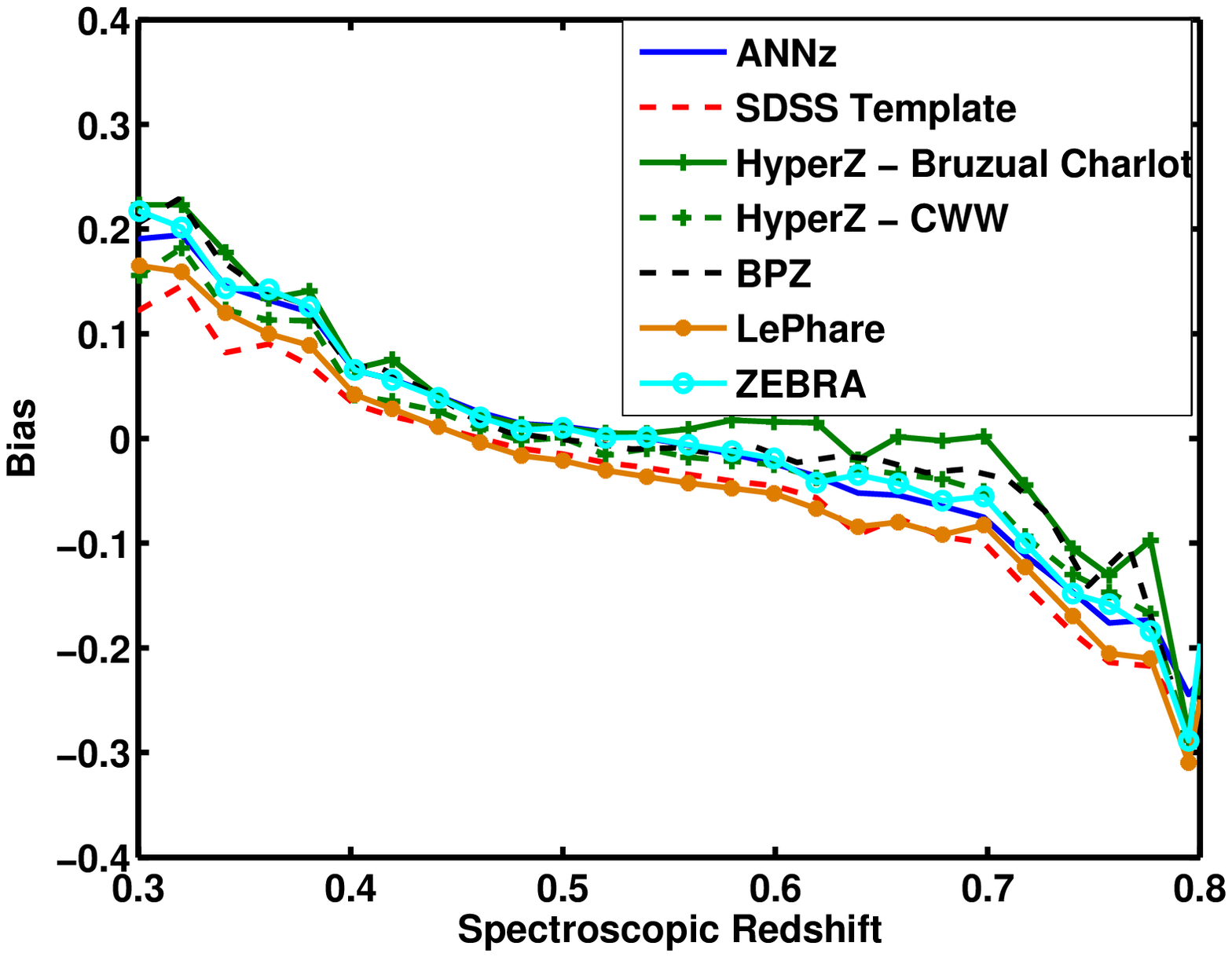}
\end{minipage}
\caption{$1\sigma$ scatter around the mean photometric redshift plotted 
as a function of spectroscopic redshift according to Eq.\ref{eq:scatter2} 
(left panel) and bias as a function 
of spectroscopic redshift (right panel). We can see that there is a similar 
trend for most codes but a difference is present. In these metrics it seems 
that the training code is better suited to the scatter but not for the 
bias. We can see from the next 2 figures that the opposite is true.}
\label{fig:scatter2}
\end{center}
\end{figure*}

\begin{figure*}
\begin{center}
\begin{minipage}[c]{1.00\textwidth}
\centering 
\includegraphics[width=7.5cm,angle=0]{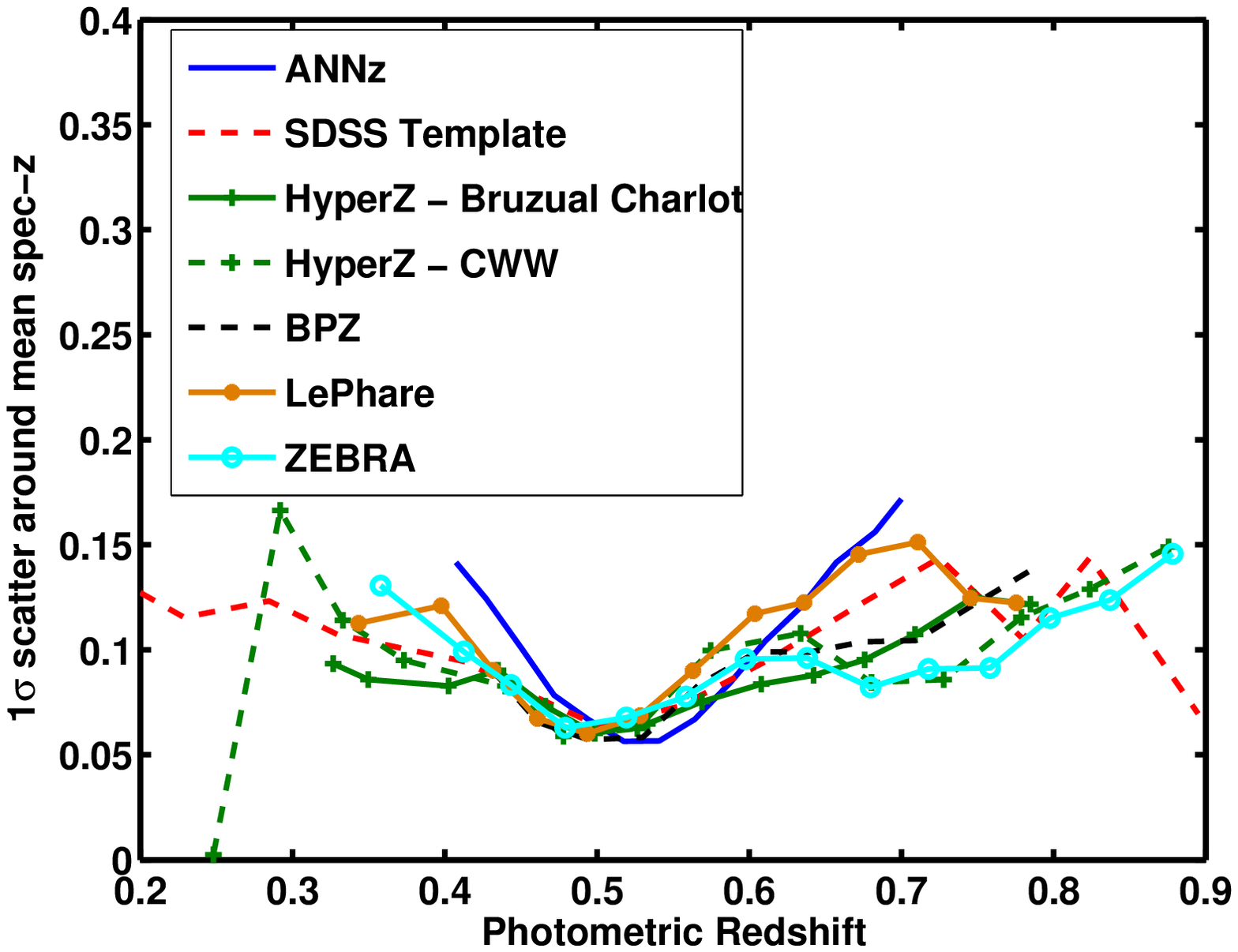}
\includegraphics[width=7.5cm,angle=0]{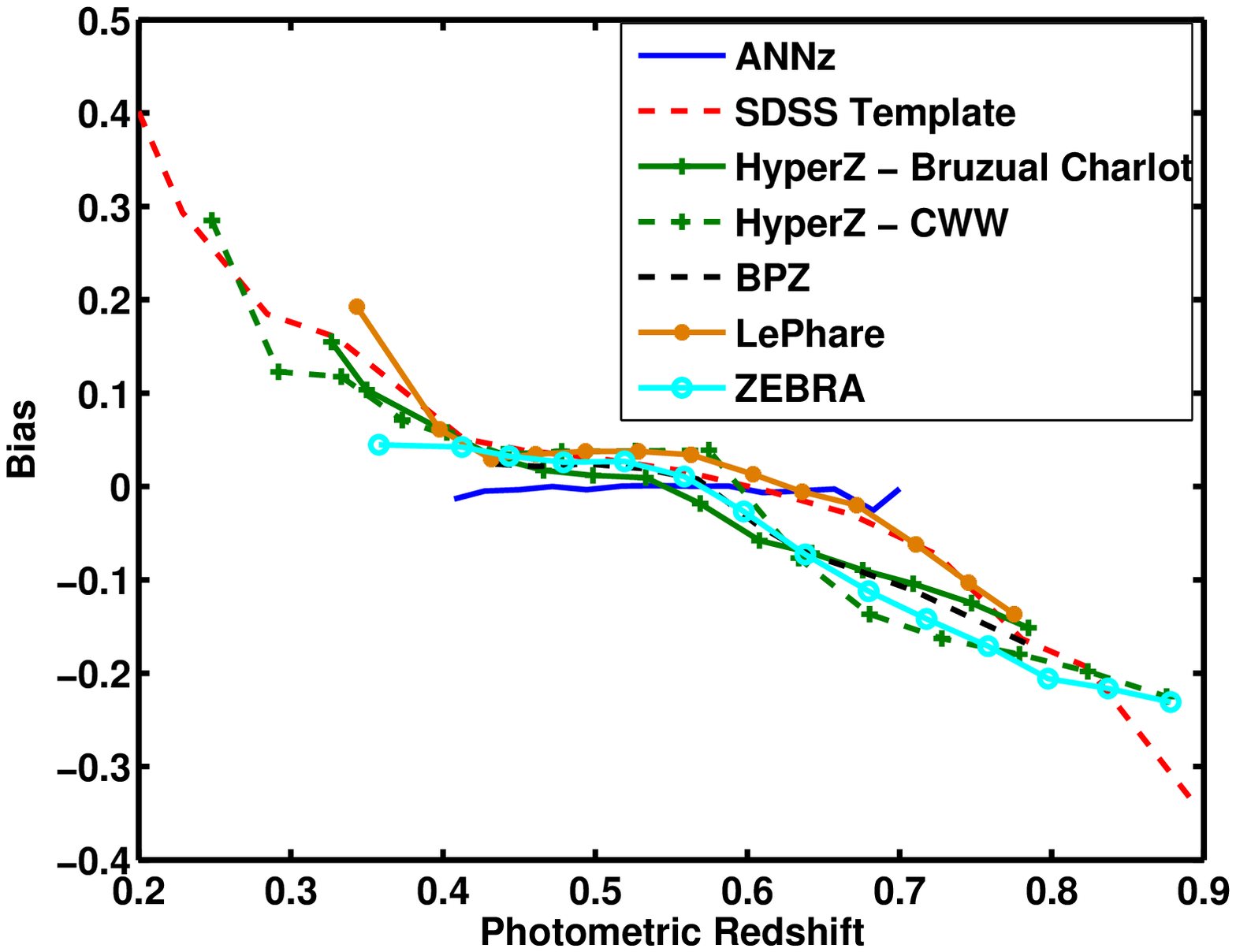}
\end{minipage}
\caption{$1\sigma$ scatter around mean spectroscopic redshift as a function of photometric redshift according to Eq.\ref{eq:scatter3} in the left panel and bias as a function of photometric redshift in the right panel. We can clearly see the power of the training code looking at the bias graph, ANNz performs with virtually no bias. However this has a certain drawback, the scatter around the mean spectroscopic redshift is larger in certain areas as a function of photometric redshift. Clearly different methods produce different quality results depending on the figure of merit used. We note e.g. that ANNz has limited coverage in photo-z as the training set is confined to that redshift range.}
\label{fig:scatter3}
\end{center}
\end{figure*}

\subsubsection{BPZ}
\label{sec:BPZ}

An extension of the above HyperZ 
likelihood ($\chi^2$) approach is to incorporate
priors, with the Bayesian framework.
\citet{2000ApJ...536..571B} formulated the problem as follows.
The probability of a galaxy with colour  $C$ and magnitude $m$
having a redshift $z$ is

\begin{equation}
 p(z | C,m) = \frac{p(z | m) p(C | z)}{p(C)} \propto p(z|m) p(C | z), \; 
\end{equation}

\noindent where the term $p(C|z)$ is the conventional redshift likelihood
employed e.g. by Hyper-z, and $p(C)$ is just a normalisation. The new
important ingredient is $p(z | m)$ , which brings in the prior
knowledge of the magnitude redshift distribution.  With the aid of the
extra information (prior), this approach is effective in avoiding 
catastrophic errors of placing a galaxy at an unrealistic redshift.

BPZ can function in a Bayesian and Maximum Likelihood 
(ML hereafter) module and 
therefore produces two outputs for the photometric redshift. 
The ML method simply picks the highest maximum over all the 
likelihoods as its redshift estimate whereas the Bayesian method 
averages over all the likelihoods after weighting them by their 
prior probabilities.

BPZ also takes as its input a photometric catalogue with magnitudes 
in different filters and their corresponding errors. 
The BPZ templates include the four CWW templates as well as 
the spectra of two star-bursting galaxies from \citet{1996ApJ...467...38K}. 
In this study we added nine interpolations between each of the four 
CWW templates to the BPZ template list to produce a more complete 
list of basis SEDs. This gives us a set of 38 basis templates for 
BPZ to use. We find that $\sim$17 of these templates concentrated 
towards the early types are sufficient to produce the best photometric 
redshift estimate. Adding more templates does not improve the 
photo-z scatter. We also used two further points of interpolation 
between each of the templates in colour space as specified by the 
INTERP parameter in BPZ. A flat prior was used throughout the calculation 
resulting in very similar results from the Bayesian and ML runs. 

The BPZ output includes two photo-z estimates from the Bayesian and 
ML runs as well as a quantity called $odds$ that is the amount of 
probability contained between $-0.12(1+z)$ and $0.12(1+z)$ around the 
Bayesian photo-z estimate. In order to select galaxies which only have 
a single compact peak in their probability distribution, we need to 
consider those galaxies with $odds>0.95$ at the very least and $odds>0.99$ 
for a robust estimate (private communication: N.Benitez). In the first 
case, we select out 4811 of the 5482 galaxies and in the second case we 
are left with 3689 of the 5482 galaxies. The Bayesian output from BPZ 
with galaxies with $odds>0.99$ selected, gives us the best photo-z 
estimate and it is this result that we use as the BPZ output in the plots.




\subsubsection{Le PHARE - PHotometric Analysis for Redshift Estimations}

Le PHARE is very similar to HyperZ in that a set of template 
SEDs together with a filter set are used to determine a set of model 
magnitudes used in the photometric redshift calculation. These are then 
compared to the observed magnitudes using a $\chi^2$ minimisation in 
order to compute the redshift of an object. The Le PHARE package 
includes various template sets used to construct the library of model 
magnitudes. These include the Coleman, Wu and Weedman and Kinney 
star-burst templates, an extended CWW template set with 72 interpolations 
between the standard CWW templates, 42 synthetic GISSEL templates as 
well as the observed templates of Poggianti. We have experimented with 
using these various template sets for photo-z estimation on our sample 
of 5482 2SLAQ objects and find the best photometric redshifts to be 
obtained with the 8 Poggianti templates corresponding to galaxy types 
Ell, S0, Sa, Sb, Sc, Sd, SB2 and SB1. The 42 GISSEL templates give 
slightly worse photo-z's than the Poggianti templates but the scatter 
on the photometric redshift when using the extended sample of 72 
interpolated CWW templates is $\sim$30\% worse than that obtained 
using the 8 Poggianti templates. Therefore we can see that in a 
template-based method, we do not necessarily gain in redshift 
accuracy by adding more model SEDs to our library. 

Le PHARE also includes various prescriptions to correct for galactic extinction. We have tried running Le PHARE with different extinction laws assuming E(B-V)=0.034 and find that in all cases, the photo-z estimate is worse when we include the effects of galactic extinction. Our final Le PHARE output is therefore obtained using the 8 Poggianti templates and neglecting the effects of galactic extinction. We use 5-band $ugriz$ photometry in the SDSS filters as this gives significantly more accurate photometric redshifts compared to if we remove the $u$-band photometry with the largest photometric errors. 

\subsubsection{ZEBRA - Zurich Extragalactic Bayesian Redshift Analyzer}

The Zurich Extragalactic Bayesian Redshift Analyzer \citep{Feldmann:ZEBRA} is a more sophisticated Bayesian template-fitting photometric redshift code compared to its predecessor, BPZ. The basic principles of estimating redshifts using templates and Bayesian priors remains as described in \S \ref{sec:BPZ} but among the novel techniques employed within the ZEBRA package are the photometry check mode that checks and corrects the photometry in certain filters, a template optimisation mode to improve the standard set of templates in specified redshift bins using a training set of galaxies with spectroscopic redshifts and the ability to calculate a prior self-consistently from the photometric catalogue when ZEBRA is run in its Bayesian mode. 

We choose not to employ the photometry check mode within ZEBRA as we are fairly confident that we have reliable photometry for our objects and have checked that applying a catalogue correction does not improve the photometric redshifts. 

ZEBRA's template set consists of the standard E, Sbc, Scd and Im galaxies as well as the SEDs of the two star-bursting galaxies SB2 and SB3. These are further interpolated in logarithmic space by ZEBRA during the photo-z estimation. We find that including the E, Sbc and Scd templates produces better results than including all six templates. Furthermore, we use ZEBRA's template optimization mode to construct improved templates from these three basis templates in two redshift bins - $0<z<0.5$ and $0.5<z<1.0$. We use a regularisation parameter of $\rho=0.05$ and a pliantness parameter of $\sigma=2$. \citet{Feldmann:ZEBRA} gives details of these parameters and how to optimise them so as to produce the most realistic templates. We do not include IGM absorption in our templates as we find that this produces better photometric redshifts. The optimisation procedure produces 39 basis templates and we use these along with our original templates in the photometric redshift calculation. 

We find the Bayesian mode of ZEBRA to produce considerably better photometric redshifts than the Maximum Likelihood mode and we therefore consider only outputs from the Bayesian mode when calculating figures of merit. The Bayesian mode is run using four iterations to calculate the prior self-consistently from the photometric catalogue. Further iterations slightly worsen the photo-z estimate. We use a smoothing kernel to smooth the prior after every iteration.

\section{Results}
\label{sec:results}

Here we summarise the results obtained when running our different public 
photo-z codes on the 2SLAQ sample of 5482 LRGs. 
Note that each of the photo-z codes detailed in 
$\S$ \ref{sec:codes} have been run several times using different 
parameters in order to optimise them to produce the best photo-z estimate. 
The final output files from each code used in this analysis are ones that 
gave the best photo-z estimate, hence this is not only a code comparison 
it is a code plus library comparison which is the final 
publicly available product to the non-expert on-line.
We note here that this is not a comparison which is meant to contrast equal values. 
For instance it is already well accepted that training codes
work much better within the redshift and spectral energy distribution 
range present in the training set but
template methods are superior if there are objects in the survey outside
this range. Also, codes such as BPZ provide an automatic selection of the objects with the best photo-z's via for example the odds parameter. 
The purpose here is to compare the full packages, including SEDs and features available from different codes. Furthermore, the chosen
sample of Luminous Red Galaxies has a very narrow range of SEDs and this comparison therefore does not highlight
the strength of photometric redshift codes with a broad range of library templates that would be more suitable for other samples. 

In Figure \ref{fig:density} we plot density plots of the 
spectroscopic redshift versus photometric redshift for 
each of these different codes. We use a redshift resolution
for all codes which ensures that the main uncertainty is related to the photo-z
uncertainty and not to numerical effects.

In order to evaluate the precision with which each of these 
different codes calculates the photometric redshift, 
we can look at the $1\sigma$ scatter between the true 
(spectroscopic) redshift and the photometric redshift. 
This is defined as follows:

\begin{equation}
\sigma_z=\left<\left(z_{phot}-z_{spec}\right)^2\right>^{\frac{1}{2}}
\label{eq:scatter1}
\end{equation}

This quantity is plotted in the left-hand panel of 
Figure \ref{fig:scatter1}. As expected, the empirical photo-z 
estimator, ANNz seems to work best at intermediate redshifts where there 
are a large number of representative training set galaxies. At high redshifts, 
HyperZ BC provides us with the best estimates 
of the photometric redshift. At low redshifts, the SDSS code and 
Le PHARE template fitting codes perform the best. We
note that none of these runs used the more standard CWW templates suggesting that
these templates are not a good match to the LRGs that are being analysed in this study.

As can be seen in Figure \ref{fig:density} however, there are many 
outliers present in our sample. Another useful quantity to consider is 
therefore $\sigma_{68}$ which is the interval in which 68\% of 
the galaxies have the smallest difference between their spectroscopic 
and photometric redshifts. This will give us some indication of the 
scatter in the photometric redshift estimate once the outliers 
have been removed and is plotted in the right-hand 
panel of Figure \ref{fig:scatter1}.

Another important quantity used to quantify how good a photo-z estimate is, is the bias defined as:

\begin{equation}
b_z=\left<z_{phot}-z_{spec}\right>
\label{eq:bias}
\end{equation}

This quantity is plotted for each of the different codes, in the 
right-hand panel of Figure \ref{fig:scatter2}. \citet{2005MNRAS.359..237P} 
show that galaxies with a given photometric redshift often have a 
systematic bias on them and this bias can therefore be added to those 
photo-z galaxies in order to correct for it. In order to get a feel for 
the error on the photometric redshift once this bias has been 
corrected for, we plot in the left-hand panel of Figure 
\ref{fig:scatter2}, the $1\sigma$ scatter around the mean 
photometric redshift estimate in each bin, defined as follows:

\begin{equation}
\sigma_{z2}=\left<\left(z_{phot}-\bar{z}_{phot}\right)^2\right>^{\frac{1}{2}}
\label{eq:scatter2}
\end{equation}

As can be seen, the scatter is now reduced for most of the codes as we are not accounting for any systematic shift that can be corrected for. The bias is largest at high redshifts for the SDSS and Le PHARE template fitting codes and these codes have the biggest improvement in the scatter at high redshifts when we take the moment around the mean rather than the true redshift. 

A more useful quantity in terms of future surveys is to plot the bias and scatter as a function of the photometric redshift as in Figure \ref{fig:scatter3}. The $1\sigma$ scatter around the mean 
spectroscopic redshift estimate in each photo-z bin is defined as follows:

\begin{equation}
\sigma_{z3}=\left<\left(z_{spec}-\bar{z}_{spec}\right)^2\right>^{\frac{1}{2}}
\label{eq:scatter3}
\end{equation}

We can also compare the right-hand panels of Figure \ref{fig:scatter2} and 
Figure \ref{fig:scatter3} to each other. We can see that the bias follows the same trend as a function of
spectroscopic redshift for all the different photo-z codes and is fairly similar for all these codes.
However, the bias as a function of the photometric redshift is very different for the different photo-z codes and more
indicative of how much the photo-z estimate has to be corrected for systematic errors. This bias is almost flat for 
the training method which has enough training set galaxies to effectively minimise the bias through the training process.
The integral under the curve is also small for ZEBRA as shown in Table \ref{tab::sigbias}  as the template optimisation technique here was able to remove 
the average bias for the entire redshift range. However, a remaining bias was found at high and low redshifts
with the values of the ZEBRA configuration parameters used in this analysis\footnote{It is possible 
that a better choice of template optimisation parameters could be found resulting in a removal of the bias at low and high redshifts.
However, as the philosophy of this paper is to perform a code comparison from the point of view of the photo-z user rather than the photo-z developer, 
and it is not obvious what this better choice of parameters would be, we choose to leave our results as they are.}.

As an alternative statistic, 
we present in Table.\ref{tab::frac} the fraction statistics for the 
methods presented in this section. 
The fractions $f_i$ are defined as the fraction of galaxies in certain
regions of the photo-z/spec-z plane. If we divide this plane in 
rectangular regions, the fraction f0
is the fraction of galaxies which is on the diagonal of this matrix. 
We could have subdivided the areas along the diagonals, this would have been a
suitable statistic as well. However there are reasons for which 
the statistic we quote is interesting. For instance if one is 
interested in cosmological probes where the galaxies are 
separated in photo-z bins then a subdivision along the photo-z axis is more natural. 
Given that these galaxies are relatively red 
with good photometric redshifts, we can see that the fraction of outliers is 
small for all estimators. however there is still a difference between different
implementations of publicly available codes.

We also present in Table.\ref{tab::sigbias} the integrated bias and scatter 
for all the codes and libraries we have used. We can see that the training
code performs best which is to be expected with a complete training set and 
that the bias here is very small. However this statistic does not
show the redshift dependence of the scatter or bias which may be of 
interest depending on the application.

\section{Systematic checks on the photometric catalogues}
\label{sec:harmonics}

The clustering of the SDSS LRG photometric sample has been analysed using the 
SDSS code for photometric redshifts described in Sec.\ref{sed::sdss_code}
by \citet{2006astro.ph..5302P}
and by ANNz described in Sec.\ref{sec::annz_code} \citet{2007MNRAS.374.1527B}. 
For the rest of this section we choose to look more in detail at the
effects that these two codes have on the end products of the analysis.
We check for gradients in the photo-z distribution across
the sky. These gradients should be present if the training set for 
the neural networks is biased as a function of position in the sky. 

\begin{table}
  \begin{center}
    \begin{tabular}{|l|c|c|c|c|c|c|c|}
      \hline       $Method$ & $f0$ & $f1$ & $f2$ & $f3$ & $f4$  \\
      \hline
      $ANNz$        &  37\% & 28\% & 31\% & 4\% & 0.02\% \\
      $SDSS$        &  27\% & 31\% & 37\% & 4.5\% & 0.8\% \\
      $HyperZ CWW$  &  24\% & 32\% & 35\% & 7\% & 1.5\% \\
      $HyperZ BC $  &  26\% & 32\% & 32\% & 7\% & 1.5\% \\
      $LePHARE$     &  26\% & 34\% & 34\% & 4\% & 0.5\% \\
      $ZEBRA$       &  26\% & 34\% & 34\% & 5\% & 0.8\% \\
      $BPZ$         &  24\% & 32\% & 36\% & 6\% & 1\% \\
      \hline
    \end{tabular}    \vspace{2mm}
  \end{center}
  \caption{Fraction statistics for the different codes presented in this section. The Fractions f0 to f4 are defined in the following way. The galaxies are
divided into a matrix defined along the axes in the photo-z/spec-z plane. The fraction f0
is the fraction of galaxies which is on the diagonal of this matrix. The fraction f1 is the fraction of galaxies in the first off diagonals of the matrix and so on. The grid are defined with the following boundaries in redshift $z = [0.2,0.4,0.5,0.6,0.7,0.9]$. We can see from the table that different methods provide
different number of outliers. Given the nature of the galaxies the outlier fraction is small in each method but still relatively different across methods.
\label{tab::frac}}
\end{table}

\begin{table}
  \begin{center}
    \begin{tabular}{|l|c|c|c}
      \hline       $Method$ & $\sigma_z$ & $bias$ \\
      \hline
      $ANNz$        &  0.0575  & 0.0014 \\
      $SDSS$        &  0.0808  & -0.0264\\
      $HyperZ CWW$  &  0.0973  & -0.0076 \\
      $HyperZ BC $  &  0.0862  & 0.0160\\
      $LePHARE$     &  0.0718  & -0.0302 \\
      $ZEBRA$       &  0.0898  & 0.0013\\
      $BPZ$         &  0.0933  & 0.0112\\
      \hline
    \end{tabular}    \vspace{2mm}
  \end{center}
  \caption{Average 1$\sigma$ scatter (Eq. \ref{eq:scatter1}) and bias (Eq. \ref{eq:bias}) for the entire sample for different methods. 
This is yet another metric to use if the redshift dependence of the bias and scatter is not of interest.
Note that the definition of the 1$\sigma$ scatter here is different from that in \citet{2007MNRAS.375...68C} Eq. 10
\label{tab::sigbias}}
\end{table}

\subsection{Checking for Gradients in Redshift Difference across the Sky}

\begin{figure}
\begin{center}
\includegraphics[width=8.0cm, angle=0]{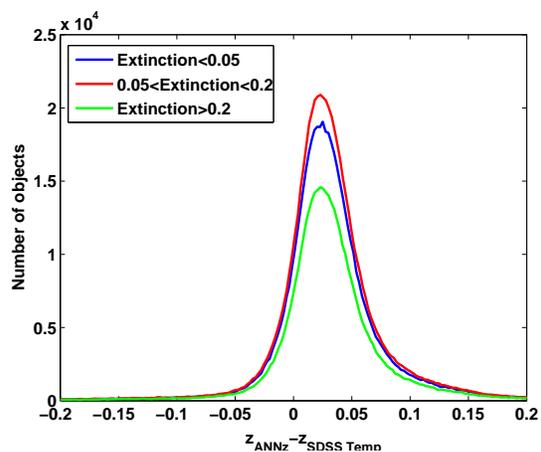}
\caption{Histograms of the difference between the photometric redshifts from the ANNz code and from the
SDSS template fitting code. We can see that the curves are identical apart from the normalization which 
is due to the different number of galaxies in each bin. this shows that the extinction
is not producing a significant bias in the photometric redshifts given that one has a 
training set which is limited in the area in the sky.\label{fig:ext}}
\end{center}
\end{figure}

\begin{figure*}
\begin{center}
\includegraphics[width=12.0cm, angle=0]{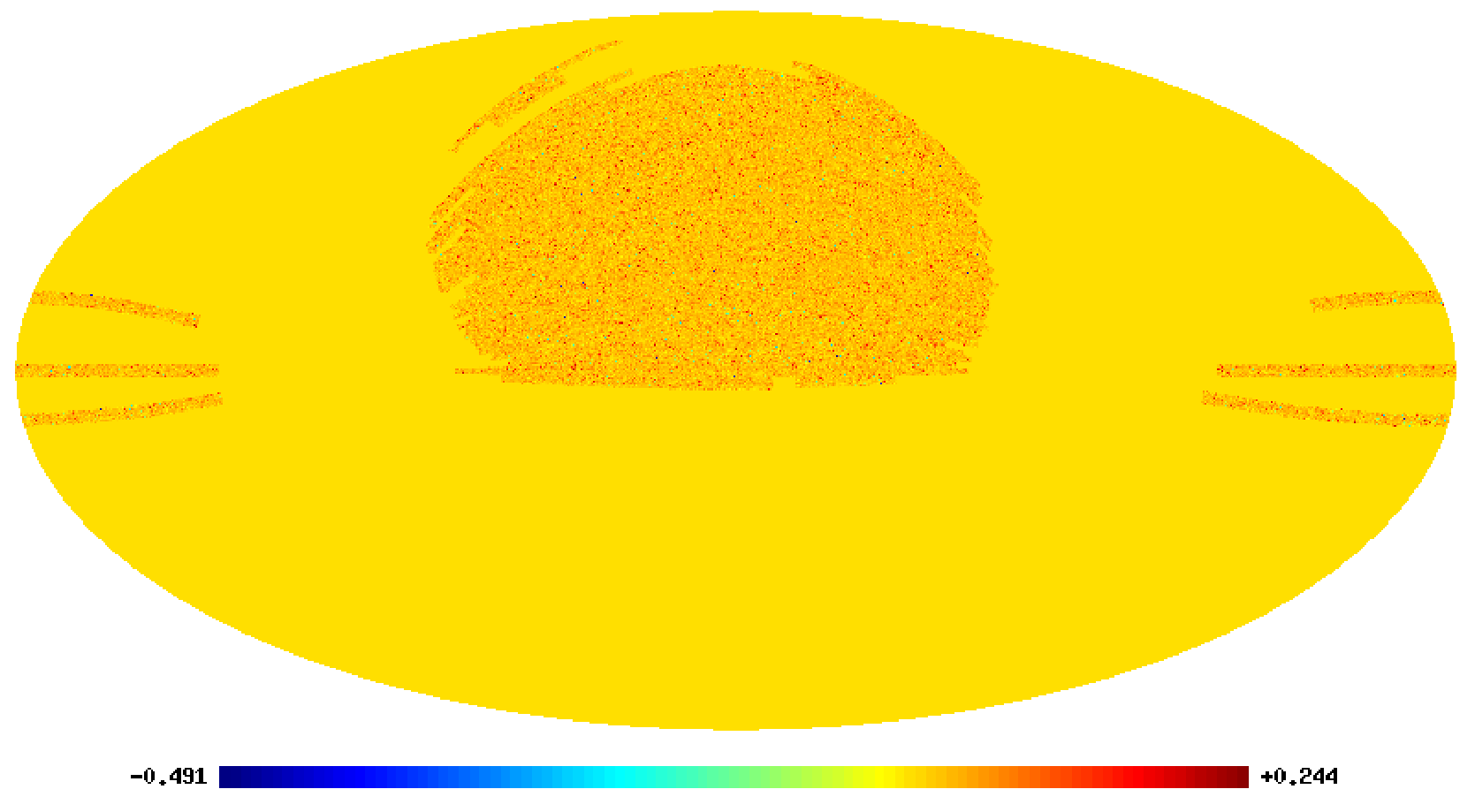}
\includegraphics[width=12.0cm, angle=0]{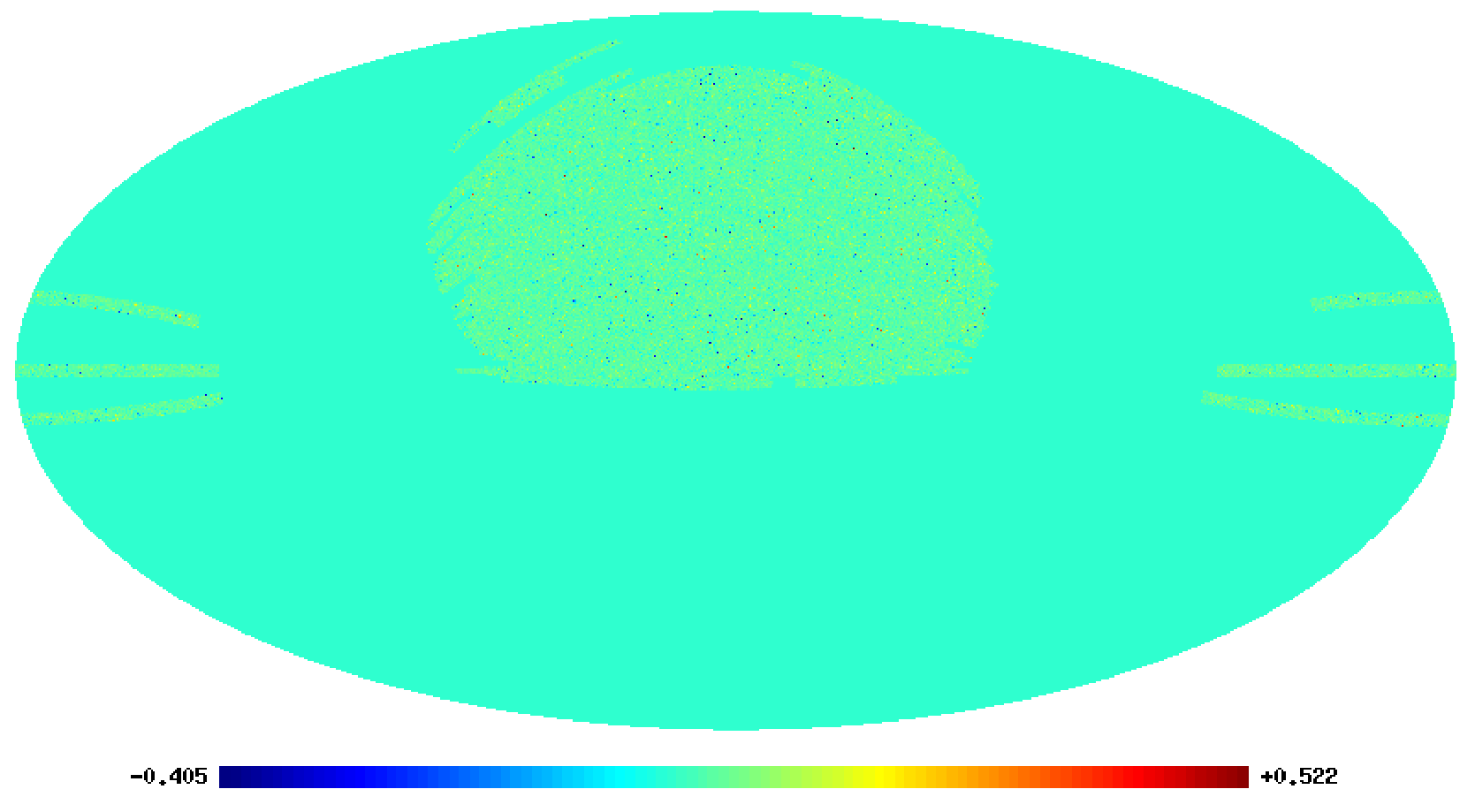}
\includegraphics[width=12.0cm, angle=0]{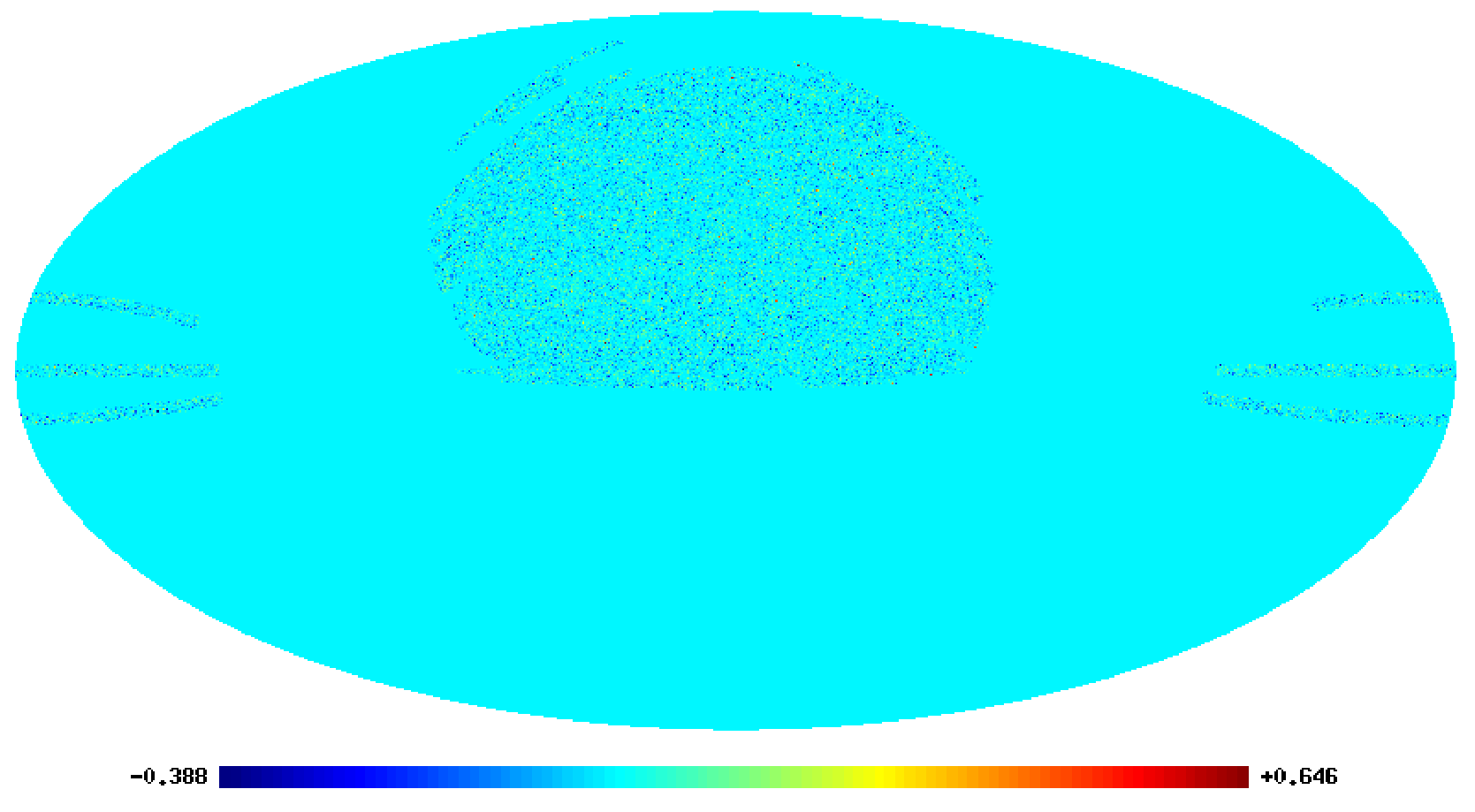}

\caption{The bias between the photo-z estimate from ANNz and the photo-z estimate
from SDSS. This has been subdivided in bins according to the photo-z estimate from ANNz.
The redshift bins are 0.4 to 0.5 (top), 0.5 to 0.6 (middle) and 0.6 to 0.7 (bottom).
We can clearly see from the colour coding that there is a bias as a function of redshift. 
If the bias is disregarded and we look at the variation across the sky there is no 
evidence that there is a gradient or that the photometric redshifts are different closer to the
regions where the training set was drawn from. The random nature of the residual bias shows that 
the extrapolation in sky position and calibrations are done to a sufficient accuracy
and that the photo-z are statistically reliable in the plane of the sky. \label{fig:sky}}
\end{center}
\end{figure*}

\begin{figure*}
\begin{center}
\includegraphics[width=15.0cm, angle=0]{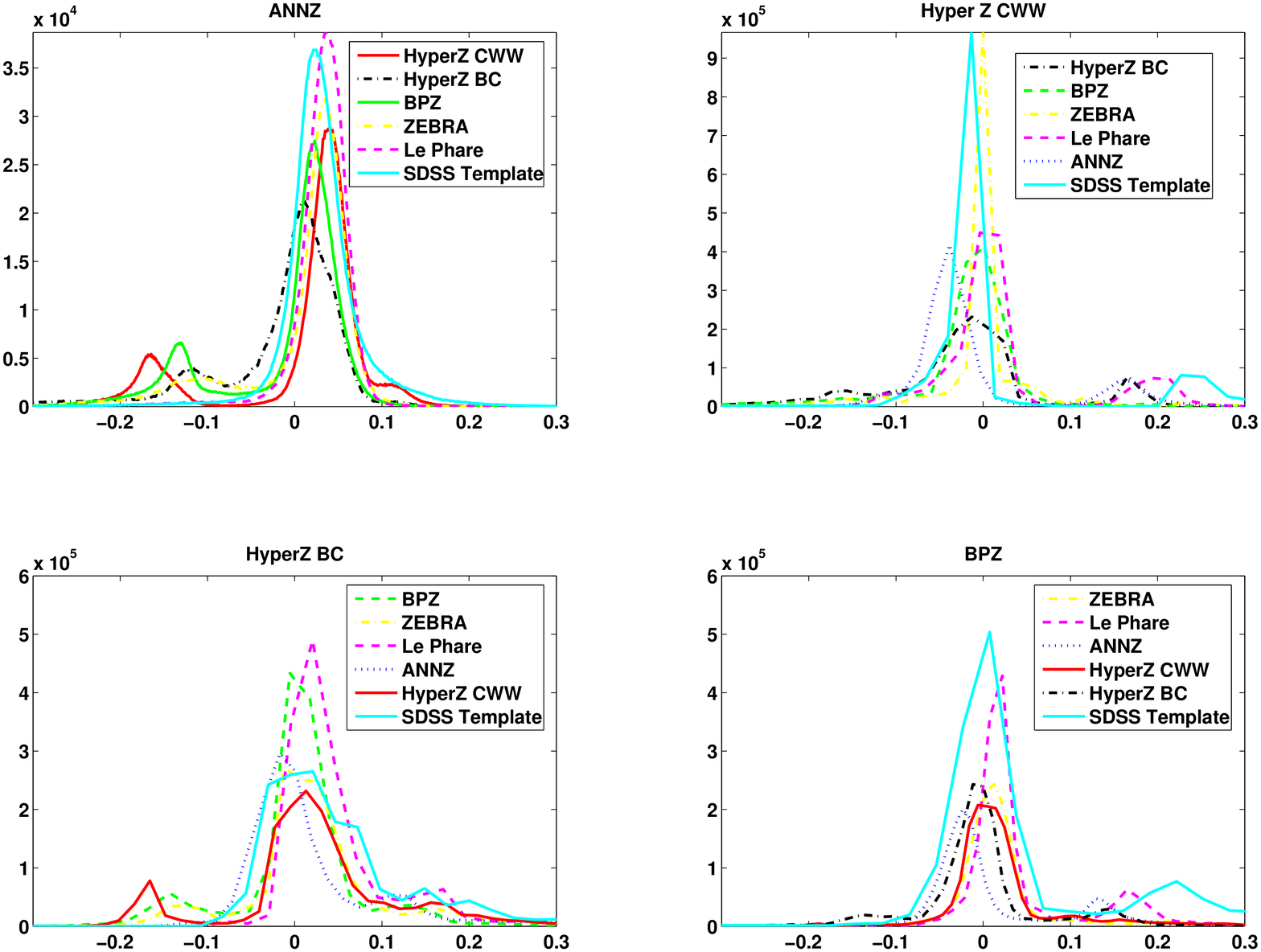}
\includegraphics[width=15.0cm, angle=0]{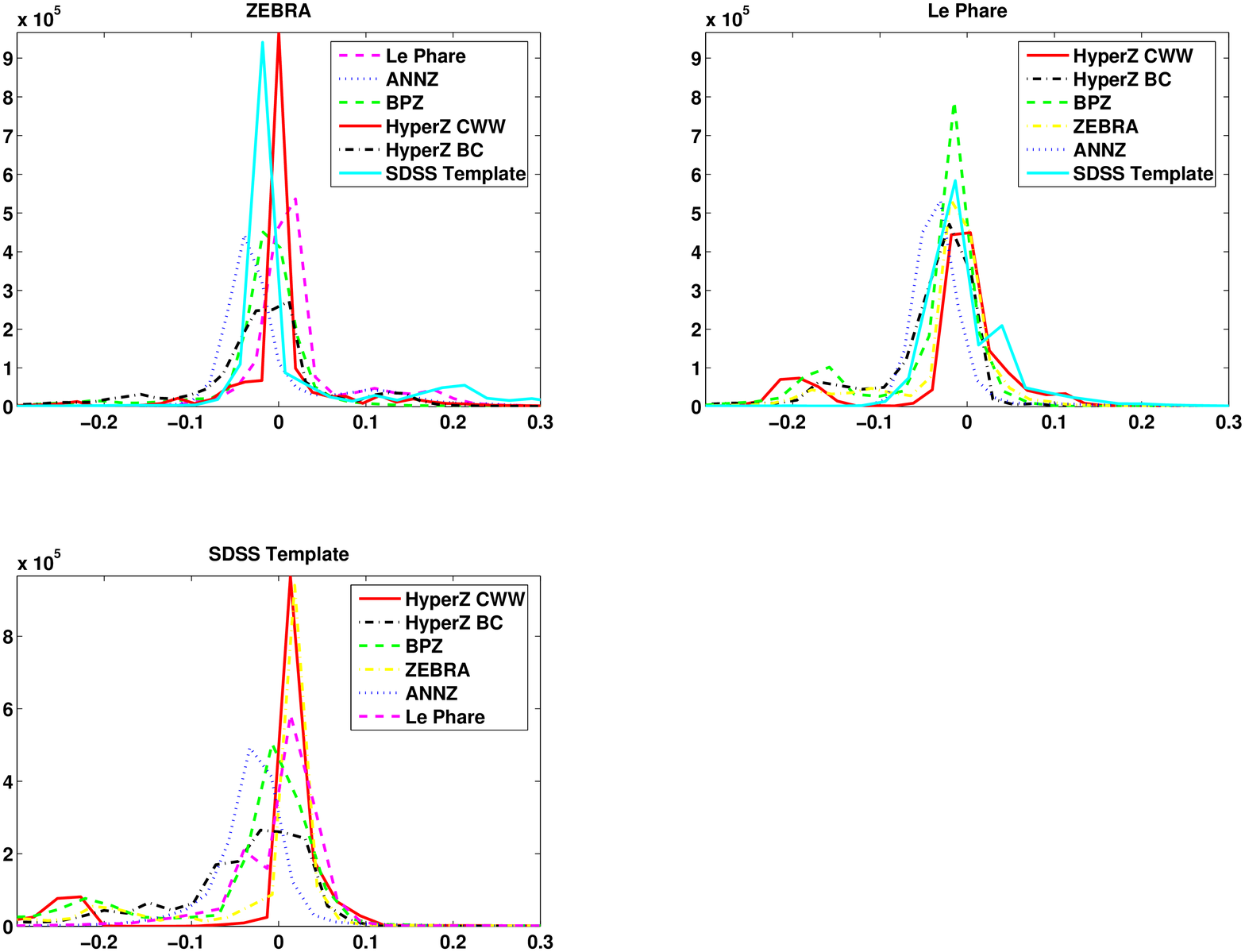}
\caption{Histogram of the difference between the
photometric redshift estimation between all pairs of codes we have used in this analysis.\label{fig:comp}}
\end{center}
\end{figure*}

It is difficult to trust two very different techniques such as template
fitting methods and training methods to produce consistent results
without making a comprehensive comparison of both methods on the same set
of galaxies.
Given that the training set from the training method is drawn from a small
region of the sky with limited range of galactic extinction one could assume a 
priori that this extra calibration which is necessary might introduce biases
as a function of sky position in applications of empirical methods. This would be
dramatic in the case of, for instance cosmological studies where
we are attempting to calculate variations across the sky to infer cosmological parameters.

We have looked for gradients in the difference $dz = z_{ANNz} - z_{SDSS}$ in 
three redshift shells, 0.4 $<$ $z_{phot}$ $<$ 0.5, 0.5 $<$ $z_{phot}$ $<$ 
0.6 and 0.6 $<$ $z_{phot}$ $<$ 0.7.
We have separated galaxies in each redshift bin according to the ANNz photometric redshift.
If the separation were done with the SDSS photometric redshifts instead the
result would not change a lot. The values of $dz$ were taken
as an average value in pixels produced with HEALPIX, 
hence are smoothed to produce the maps in Fig.\ref{fig:sky}. 
No apparent gradients can be identified, in any of the redshift shells, 
which is an indicator for the consistency across the plane of the sky.

\begin{table*}
\caption{Parameters included in the updated MegaZ-LRG DR6 photometric redshift catalogue.}
\label{tab.catpars}
\centering
\begin{tabular}{ll}
\hline 
\texttt{objID} & SDSS objID\\
\texttt{ra} &J2000 right ascension \\
\texttt{dec} &J2000 declination\\
\texttt{dered\_u}&\\ 
\texttt{dered\_g}&\\
\texttt{dered\_r} &Dereddened model magnitudes\\
\texttt{dered\_i} &\\
\texttt{dered\_z} &\\
\texttt{err\_u}&\\ 
\texttt{err\_g}&\\
\texttt{err\_r} &Magnitude errors\\
\texttt{err\_i} &\\
\texttt{err\_z} &\\
\texttt{deVMag\_i} &Dereddened de Vaucouleurs magnitude\\
\texttt{z\_annz}     & ANNz photometric redshift\\
\texttt{z\_annz\_err} & ANNz photometric redshift error\\
\texttt{delta\_sg}& ANNz galaxy probability\\
\texttt{delta\_err\_sg}& ANNz galaxy probability error\\

\texttt{z\_sdss}     & SDSS photometric redshift\\
\texttt{z\_hzcww}     & Hyper-z CWW photometric redshift\\
\texttt{z\_hzcww\_chi} & Hyper-z CWW chi squared\\
\texttt{z\_hzcww\_errl} & Hyper-z CWW photometric redshift 68\% lower confidence limit\\
\texttt{z\_hzcww\_errh} & Hyper-z CWW photometric redshift 68\% higher confidence limit\\
\texttt{z\_hzbc}     & Hyper-z BC photometric redshift\\
\texttt{z\_hzbc\_chi} & Hyper-z BC chi squared\\
\texttt{z\_hzbc\_errl} & Hyper-z BC photometric redshift 68\% lower confidence limit\\
\texttt{z\_hzbc\_errh} & Hyper-z BC photometric redshift 68\% higher confidence limit\\
\texttt{z\_bpz\_bayes}     & BPz bayesian photometric redshift\\
\texttt{z\_bpz\_errl} & BPz photometric redshift 90\% lower confidence limit\\
\texttt{z\_bpz\_errh} & BPz photometric redshift 90\% higher confidence limit\\
\texttt{z\_bpz\_odds}     & BPz bayesian odds parameter\\
\texttt{z\_bpz\_ml}     & BPz maximum likelihood photometric redshift\\
\texttt{z\_bpz\_chi} & BPz chi squared\\
\texttt{z\_zebra}      & ZEBRA photometric redshift\\
\texttt{z\_zebra\_errl} & ZEBRA photometric redshift 68\% lower confidence limit\\
\texttt{z\_zebra\_errh} & ZEBRA photometric redshift 68\% higher confidence limit\\
\texttt{z\_lp}      & Le PHARE photometric redshift\\
\texttt{z\_lp\_prob} & Le PHARE percentage PDF between $dz=z_{best}\pm0.1(1+z_{best})$\\
\hline
\end{tabular}
\centering
\end{table*}

The different colour coding in each of the redshift shells in Fig.\ref{fig:sky}
are indicative of the bias between the two methods which is of course still present,
however taking that bias aside, there seems to be no correlation between the 
usual regions of high extinction in the SDSS regions and the scatter of biases as a function
of sky position produced here. Given that the template set does not know about the
training set which belongs to a selective region of the sky our conclusion is that
the fact that the training set is restricted to a small region of the sky 
does not include significant biases as a function of sky position and therefore 
is not an extra source of systematic biases. This could be taken further, for instance by calculating 
spherical harmonics of the map above and comparing with theoretical predictions in order to 
estimate the actual lower bound of a potential systematic effect for a given probe
but since this would involve a more specific cosmological approach we argue that this
is beyond the photometric redshift comparison which is the aim of this paper.

The main reason why there should be a systematic bias as a function of sky position 
is extinction. We have computed the difference between the two photo-z estimates for 
different regions of galactic extinction and plotted histograms for these quantities in Fig.\ref{fig:ext}.
Apart from the normalisation which encoded the fact that there are a different number of galaxies in
these bins the curves are virtually identical with same bias and scatter, this also shows that there
is no evidence of significant differences. This can only get better with future Planck data. 
So we are confident that this is not a systematic effect that will hinder future or current photometric redshift 
analysis.

\subsection{A Photo-z comparison of the codes}

We also present here a comparison of how the photometric redshifts of each code compare to each other.
We present in Fig.\ref{fig:comp} histograms of the difference between
photo-z for each pair of codes that we have used in the analysis of our updated MegaZ-LRG catalogue.
We can see the differences between codes is apparent in some plots.

For instance comparing ANNz to other codes there seem to be some outliers at 
a redshift difference of 0.1 compared to Hyperz CWW, BPZ and Hyperz BC. 
Similarly other pairs of codes produce outliers which indicates that this is not
only a difference between template codes and training codes.
We also note that for instance comparing ZEBRA with codes such as HyperZ CWW
or SDSS code there is a good agreement on the scatter but there is a small bias between the codes which
may suggest that the templates used might not have been optimal in some codes. The ImpZ code (private communication:M. Rowan-Robinson)
was also tested on the sample of 5482 2SLAQ LRGs and produced consistent results with other template-based methods. However, as the code in
its current form is not yet publicly available, we do not present these results here or extend the analysis to the MegaZ-LRG DR6 catalogue described in the next section. 

We emphasise here that there are many differences even 
though all the photo-z estimates are of relatively good quality.
There is therefore a need to deconstruct the effects of the algorithm and the template libraries 
in order for us to understand these differences and have even more reliable photo-z's in the future.

\section{An extension to MegaZ-LRG: catalogues with different photo-z estimators for SDSS DR-6}
\label{sec:catalogue} 

We have extended the photometric sample from SDSS DR4 to SDSS DR6 imaging 
catalogue using the same criteria devised for the 2SLAQ LRG catalogue.
This extended MegaZ-LRG catalogue contains 1543596 objects over more than 8000
square degrees of the sky. As in previous studies LRGs 
are expected to be about 95\% of the sample and M type stars are 
expected to be 5\% of the sample. We have produced photometric redshift 
results for 7 different photometric redshift estimators
and provided the error estimators associated with each method. 
We also provide trained empirical values to perform star/galaxy 
separation based on a set of 15 photometric parameters as in \citep{2007MNRAS.375...68C}. 
All the parameters included in the revised catalogue are described in Table.\ref{tab.catpars}. The data can be found in the following website\footnote{www.star.ucl.ac.uk/$\sim$mbanerji/MegaZLRGDR6/megaz.html}.

\section{Conclusions}
\label{sec:concl}

We have presented an updated version of the MegaZ-LRG 
catalogue. This catalogue
contains about 1.5 million objects with accurate 
photometric redshifts which can be used for a 
range of science applications. The catalogue is available on-line 
and contains SDSS ID information
so all SDSS data can be retrieved for 
each object as well as the photometric redshifts from each of the six public codes.

We have run several comparisons of code and template libraries on the 2SLAQ LRG sample. 
We conclude that there are differences in the 
codes and stress that a more thorough comparison is
needed where the effects of the codes and template libraries are disentangled. This will allow us to pinpoint where the discrepancies are arising. 
An approach based on first principles such as that presented in \citet{2008arXiv0811.2600B} is also timely. 
We have used several figures of merit to assess 
which code + template library performed best for this set of galaxies.
We conclude that different codes perform with different 
strengths depending on the figure of merit used. We outline more 
specifically the findings below:

\begin{itemize}

\item As expected, the availability of a complete training set means the
training method, ANNz performs best in the intermediate redshift bins 
where there are plenty of spectroscopic redshifts.

\item Le PHARE performs very well particularly in the lower redshift bins suggesting the Poggianti templates may be a better fit to LRGs at those redshifts
compared to other templates used in this comparison.


\item  HyperZ run with Bruzual \& Charlot templates gives better results than using the same with CWW templates once again highlighting the importance of template choice.

\item The SDSS template code gives very good results compared to other codes at the highest photo-z bins despite having only one evolving template for the LRGs. Given the narrow range in SEDs of our sample of Luminous Red Galaxies, the strengths of template-based codes with extensive template libraries are not adequately highlighted by this comparison. 

\item ZEBRA shows a small average bias indicating the importance of the template optimisation technique in removing biases.

\end{itemize}

As expected the training code performs best where 
the training set is large and complete and 
the template sets overtake the 
training code if the training set starts to 
become sparse. The importance of template
choice is highlighted by the fact that most figures of merit show
codes used in conjunction with the CWW templates to perform worse
than those using other training or synthetic templates. This suggests
that the CWW templates are not a very good match to the SEDs of these LRGs.

There is a discrepancy between the scatters found for these codes 
ranging from 0.057 to 0.097. Both values are considered good results 
for photo-z estimates 
as one would expect from LRGs but there is a clear 
difference between the different code and template combinations that are run. Given that these differences will also depend on 
galaxy type and training set size, it is imperative that we carry out a more thorough comparison
where the effects of codes and templates are deconstructed, in order to understand what factors affect the
photo-z accuracy. We caution the reader that the results presented here are
specific to a sub-sample of galaxies, namely LRGs, which have a narrow range in SEDs and a complete
and representative spectroscopic training set available.
The conclusions presented here could and probably would change if the comparison were made with 
different galaxies or a different training set size. 

We have also produced a set of tests to assess whether 
the fact that the training sets are from a restricted area in the sky affects 
the photometric redshifts significantly. We conclude 
that there is little or no difference between the results
from template methods and training set methods across the 
sky and that the difference found is not likely to be 
a source of the training set being restricted in area. 
This is promising for future wide-field photometric redshift surveys such as the Dark Energy Survey, PanStarrs, Euclid and JDEM.

\section*{Acknowledgments}

We acknowledge 
Stephane Arnouts, Tom Babbedge, Narciso Benitez, Micol Bolzonella,
Marcella Carollo,
Nikhil Padmanabhan and Michael Rowan-Robinson for useful 
comments regarding the publicly available codes 
and also for checking that our usage of the codes was acceptable.
FBA acknowledges the support of a Leverhulme Early Careers Fellowship. MB is supported by an STFC studentship.
OL would like to acknowledge the Royal Society Wolfson Research Merit Award.  

\bibliographystyle{./reference/mn2e.bst}
 
\bibliography{./reference/aamnem99,./reference/ref_data_base}

\end{document}